\documentstyle[12pt]{article}
\begin{document}

\title{On the compatible weakly-nonlocal Poisson brackets
of Hydrodynamic Type.}

\author{Andrei Ya. Maltsev}

\date{
\centerline{ L.D.Landau Institute for Theoretical Physics,}
\centerline{117940 ul. Kosygina 2, Moscow, maltsev@itp.ac.ru} 
\centerline{ University of Maryland, College Park MD 20742,USA}   
\centerline{maltsev@math.umd.edu}
}

\maketitle

\begin{abstract}
 We consider the pairs of general weakly non-local Poisson brackets 
of Hydrodynamic Type (Ferapontov brackets) and the corresponding
integrable hierarchies. We show that under the requirement of
non-degeneracy of the corresponding "first" pseudo-Riemannian metric
$g_{(0)}^{\nu\mu}$ and also some non-degeneracy requirement for the
nonlocal part it is possible to introduce a "canonical" set of 
"integrable hierarchies" based on the Casimirs, Momentum functional 
and some "Canonical Hamiltonian functions". We prove also that all 
the "Higher" "positive" Hamiltonian operators  and the "negative"
symplectic forms have the weakly non-local form in this case. The 
same result is also true for "negative" Hamiltonian operators and
"positive" Symplectic structures in the case when both 
pseudo-Riemannian metrics $g_{(0)}^{\nu\mu}$ and $g_{(1)}^{\nu\mu}$
are non-degenerate.
\end{abstract}

\centerline{\bf Introduction}

 We discuss in this paper the Poisson pencils of
weakly nonlocal Poisson brackets of Hydrodynamic Type
(Ferapontov brackets). This means that we consider a pair
of Hamiltonian operators:

$${\hat J}^{\nu\mu}_{(0)} = g^{\nu\mu}_{(0)}(U){d \over dX} +
b^{\nu\mu}_{(0)\eta}(U) U^{\eta}_{X} +
\sum_{k=1}^{g_{0}} e_{(0)k} w^{\nu}_{(0)k\eta}(U) U^{\eta}_{X}
D^{-1} w^{\mu}_{(0)k\zeta}(U) U^{\zeta}_{X}$$

$${\hat J}^{\nu\mu}_{(1)} = g^{\nu\mu}_{(1)}(U){d \over dX} +
b^{\nu\mu}_{(1)\eta}(U) U^{\eta}_{X} +
\sum_{k=1}^{g_{1}} e_{(1)k} w^{\nu}_{(1)k\eta}(U) U^{\eta}_{X}
D^{-1} w^{\mu}_{(1)k\zeta}(U) U^{\zeta}_{X}$$
where $e_{(0)k}, e_{(1)k} = \pm 1$ and $D^{-1} = (d/dX)^{-1}$
defined in a "skew-symmetric" way:

$$D^{-1} = {1 \over 2} \left[ \int_{-\infty}^{X} dX -
\int_{X}^{+\infty} dX \right] $$

and require that the expression

$${\hat J}^{\nu\mu}_{\lambda} = {\hat J}^{\nu\mu}_{(0)} +
\lambda {\hat J}^{\nu\mu}_{(1)} $$
defines the Poisson bracket satisfying Jacobi identity for any 
$\lambda$.

 Let us mention here that the brackets of this kind are
the generalization of Dubrovin - Novikov local homogeneous
brackets of Hydrodynamic Type (\cite{dubrnov1}-\cite{dubrnov3}):

\begin{equation}
\label{dnbrack}
\{U^{\nu}(X), U^{\mu}(Y)\} = g^{\nu\mu}(U) \delta^{\prime}(X-Y)
+b^{\nu\mu}_{\eta}(U) U^{\eta}_{X} \delta (X-Y)
\end{equation}

with the Hamiltonian operator:

$${\hat J}^{\nu\mu}_{DN} = g^{\nu\mu}(U) {d \over dX} +
b^{\nu\mu}_{\eta}(U) U^{\eta}_{X} $$

\vspace{0.5cm}

{\bf Theorem.} (B.A.Dubrovin, S.P.Novikov)

{\it Consider the bracket (\ref{dnbrack}) with non-degenerate
tensor $g^{\nu\mu}(U)$. From the Leibnitz identity it follows
that $g^{\nu\mu}(U)$ and 
$\Gamma^{\mu}_{\nu\eta}(U) = - g_{\nu\xi}(U) b^{\xi\mu}_{\eta}(U)$
$(g_{\nu\xi}g^{\xi\mu} = \delta^{\mu}_{\nu})$ transform as a
metric with upper indices and the Christoffel symbols
under the pointwise coordinate transformations
${\tilde U}^{\nu} = {\tilde U}^{\nu}(U)$.

 The bracket (\ref{dnbrack}) is skew-symmetric if and only if
$g^{\nu\mu}$ is symmetric and the connection $\Gamma^{\mu}_{\nu\eta}$
is compatible with the metric: $\nabla_{\eta} g^{\nu\mu} \equiv 0$.

 The bracket (\ref{dnbrack}) satisfies the Jacobi identity if and only 
if the connection $\Gamma^{\mu}_{\nu\eta}$ is symmetric and has
zero curvature: $R^{\nu\mu}_{\eta\xi} \equiv 0$. }

\vspace{0.5cm}

 It follows from Dubrovin - Novikov theorem that any bracket
(\ref{dnbrack}) with non-degenerate $g^{\nu\mu}$ can be written
locally in the "constant form":

\begin{equation}
\label{dncan}
\{n^{\nu}(X), n^{\mu}(Y)\} = \epsilon^{\nu} \delta^{\nu\mu}
\delta^{\prime}(X-Y) \,\, , \,\,\,\,\, \epsilon^{\nu} = \pm 1
\end{equation}
in the flat coordinates $n^{\nu} = n^{\nu}(U)$.

 The functionals

$$N^{\nu} = \int_{-\infty}^{+\infty} n^{\nu}(X) dX$$
are Casimirs of the bracket (\ref{dnbrack}) and the functional

$$P = \int_{-\infty}^{+\infty} {1 \over 2} \sum_{\nu=1}^{N}
\epsilon^{\nu} n^{\nu}(X) n^{\nu}(X) dX $$
is a momentum operator generating the flow 
$U^{\nu}_{T} = U^{\nu}_{X}$. The form (\ref{dncan}) can be 
considered as the Canonical form for the DN-bracket (\ref{dnbrack})   
with the non-degenerate tensor $g^{\nu\mu}$.

 It can be seen also that any functional of "Hydrodynamic Type"

$$H = \int_{-\infty}^{+\infty} h(U) dX$$
generates a "Hydrodynamic Type System"

\begin{equation}
\label{htsys}
U^{\nu}_{T} = V^{\nu}_{\mu}(U) U^{\mu}_{X}
\end{equation}
according to bracket (\ref{dnbrack}).

 Let us mention also that the brackets (\ref{dnbrack}) with
degenerate tensor $g^{\nu\mu}(U)$ of constant rank has more
complicated but also nice differential geometric structure
(see \cite{grinberg}).

 The first generalization of DN-bracket to the weakly nonlocal
case was the Mokhov-Ferapontov bracket
\cite{mohfer1}:

\begin{equation}
\label{mferbr}
\{U^{\nu}(X), U^{\mu}(Y)\} = g^{\nu\mu}(U) \delta^{\prime}(X-Y)
+ b^{\nu\mu}_{\eta}(U) U^{\eta}_{X} \delta (X-Y) +
c U^{\nu}_{X} \nu (X-Y) U^{\mu}_{Y}
\end{equation}
where $\nu (X-Y) = 1/2 \, sgn(X-Y)$, corresponding to the Hamiltonian
operator:

$${\hat J}^{\nu\mu}_{DN} = g^{\nu\mu}(U) {d \over dX} +
b^{\nu\mu}_{\eta}(U) U^{\eta}_{X} + c U^{\nu}_{X} D^{-1} U^{\mu}_{X}$$ 

\vspace{0.5cm}

{\bf Theorem.} (O.I.Mokhov, E.V.Ferapontov).

{\it Consider the bracket (\ref{mferbr}) with non-degenerate tensor
$g^{\nu\mu}(U)$. Then:

 1) The bracket (\ref{mferbr}) is skew-symmetric and satisfies 
Leibnitz  identity if and only if tensor $g^{\nu\mu}(U)$ is a metric 
with upper indices and 
$\Gamma^{\mu}_{\nu\eta} = - g_{\nu\xi} b^{\xi\mu}_{\eta}$ are the
connection coefficients compatible with  $g^{\nu\mu}(U)$.

 2) The bracket (\ref{mferbr}) satisfies the Jacobi identity if
and only if the connection 
$\Gamma^{\mu}_{\nu\eta}$ is symmetric and has the constant
curvature equal to $c$, i.e.

$$R^{\nu\tau}_{\mu\eta} = c \left( 
\delta^{\nu}_{\mu} \delta^{\tau}_{\eta} - 
\delta^{\tau}_{\mu} \delta^{\nu}_{\eta} \right) $$ }

\vspace{0.5cm}

 The bracket (\ref{mferbr}) has a weakly nonlocal form. 
However, any local translationally invariant functional

$$H = \int h(U) dX $$
generates a local system of Hydrodynamic Type with respect to
(\ref{mferbr}). Indeed, we have

$$U^{\mu}_{X} {\partial h \over \partial U^{\mu}} \equiv
\partial_{X} h$$
if $h$ does not depend on $X$ explicitly, so the application of
$D^{-1}$ gives the local expression for the corresponding flow.   

 The Canonical form of the bracket (\ref{mferbr}) was presented first
by M.V.Pavlov in \cite{pavlov1} and can be written as:

$$\{n^{\nu}(X), n^{\mu}(Y)\} = $$

\begin{equation}
\label{canmfer}
= \left( \epsilon^{\nu} \delta^{\nu\mu}
- c n^{\nu} n^{\mu} \right) \delta^{\prime}(X-Y) - 
c n^{\nu}_{X} n^{\mu} \delta (X-Y) + 
c n^{\nu}_{X} \nu (X-Y) n^{\mu}_{Y}
\end{equation}
where $n^{\nu} = n^{\nu}(U)$ are the annihilators for the bracket
(\ref{mferbr}) (on the space of rapidly decreasing functions
$n^{\nu}(X)$ at $X \rightarrow \pm \infty$ ). Also the implicit
expression for the density of $P$ was represented in \cite{pavlov1}.
 
 We will see however that the Casimirs and the Momentum operator
for the bracket (\ref{mferbr}) depend actually on the boundary
conditions imposed on the functions $U^{\nu}(X)$ for
$X \rightarrow \pm \infty$ (\cite{malnov00}) (the condition 
$U^{\nu} \rightarrow 0$, $X \rightarrow \pm \infty$ in general
is not invariant under the pointwise transformations
${\tilde U}^{\nu} = {\tilde U}^{\nu}(U)$). As was pointed out in
\cite{malnov00} we can not speak about the Casimirs and 
Momentum functional until we fix the boundary conditions at
infinity and in the general case it is better to speak about
the invariant set of $N+1$ (for MF-bracket) functionals playing
the role of either Casimirs of Momentum operator according
to the boundary conditions. Let us consider this later for
the case of more general Ferapontov brackets.

 The general Ferapontov bracket (\cite{fer1}-\cite{fer4}) has the 
form:

$$\{U^{\nu}(X), U^{\mu}(Y)\} = g^{\nu\mu}(U) \delta^{\prime}(X-Y)
+ b^{\nu\mu}_{\eta}(U) U^{\eta}_{X} \delta (X-Y) + $$
\begin{equation}
\label{gferbr}  
+ \sum_{k=1}^{g} e_{k} w^{\nu}_{k\eta}(U) U^{\eta}_{X} \nu (X-Y)
w^{\mu}_{k\zeta}(U) U^{\zeta}_{Y}
\end{equation}
$e_{k} = \pm 1$, which corresponds to the weakly nonlocal
Hamiltonian operator:

\begin{equation}
\label{gferop}
{\hat J}^{\nu\mu}_{F} = g^{\nu\mu}(U) {d \over dX} +
b^{\nu\mu}_{\eta}(U) U^{\eta}_{X} +
\sum_{k=1}^{g} e_{k} w^{\nu}_{k\eta}(U) U^{\eta}_{X} D^{-1}
w^{\mu}_{k\zeta}(U) U^{\zeta}_{X} 
\end{equation}

\vspace{0.5cm}

{\bf Theorem.} (E.V.Ferapontov).

{\it Consider the bracket (\ref{gferbr}) with non-degenerate
tensor $g^{\nu\mu}(U)$. Then:

 1) The bracket (\ref{gferbr}) is skew-symmetric and satisfies
Leibnitz  identity if and only if tensor $g^{\nu\mu}(U)$ is a metric
with upper indices and
$\Gamma^{\mu}_{\nu\eta} = - g_{\nu\xi} b^{\xi\mu}_{\eta}$ are the
connection coefficients compatible with  $g^{\nu\mu}(U)$.

 2) The bracket (\ref{mferbr}) satisfies the Jacobi identity if
and only if the connection
$\Gamma^{\mu}_{\nu\eta}$ is symmetric and the metric 
$g^{\nu\mu}$ and tensors $w^{\nu}_{k\eta}(U)$ satisfy the equations:

$$g^{\nu\tau} w^{\mu}_{k\tau} = g^{\mu\tau} w^{\nu}_{k\tau} \,\, ,
\,\,\,\,\, \nabla_{\nu}w^{\mu}_{k\eta} =
\nabla_{\eta}w^{\mu}_{k\nu} $$

$$R^{\nu\tau}_{\mu\eta} = \sum_{k=1}^{g} e_{k} \left(
w^{\nu}_{k\mu} w^{\tau}_{k\eta} - w^{\tau}_{k\mu} w^{\nu}_{k\eta}
\right) $$

 Moreover, this set is commutative $[w_{k},w_{k^{\prime}}] = 0$. }

\vspace{0.5cm}

 It was pointed out by E.V.Ferapontov that the equations written 
above are Gauss and Petersson-Codazzi equations for the submanifold
${\cal M}^{N}$ with flat normal connection in the pseudo-Euclidean
space $E^{N+g}$. In this consideration tensor $g^{\nu\mu}$ is
the first quadratic form of ${\cal M}^{N}$ and $w^{\nu}_{k\eta}$
are the Weingarten operators corresponding to $g$ parallel 
vector fields in the normal bundle ${\bf N}_{k}$, such that
$\langle {\bf N}_{k}, {\bf N}_{m}\rangle = e_{k} \delta_{km}$.
It was also proved by E.V.Ferapontov that these brackets can be
constructed as a Dirac restriction of the local DN-bracket

$$\{Z^{I}(X), Z^{J}(Y)\} = \epsilon^{I} \delta^{IJ}
\delta^{\prime}(X-Y) \,\, , \,\,\, I,J = 1,\dots, N+g, \,\,\,
\epsilon^{I} = \pm 1 $$
in $E^{N+g}$ to the submanifold ${\cal M}^{N}$ 
(\cite{fer2},\cite{fer4}).

 As far as we know the cases of brackets (\ref{mferbr}),
(\ref{gferbr}) with the degenerate tensors $g^{\nu\mu}(U)$ were
not studied in the literature.

 All the brackets (\ref{dnbrack}), (\ref{mferbr}) and (\ref{gferbr})
are closely connected with the diagonalizable integrable systems
(\ref{htsys}). 

 The general procedure of integration of the
so-called "semi-Hamiltonian" diagonal systems of Hydrodynamic Type
was constructed by S.P.Tsarev (\cite{tsarev1}, \cite{tsarev2}). 
It can be shown that any diagonal system (\ref{htsys}) which is
 Hamiltonian with respect to bracket (\ref{dnbrack}), (\ref{mferbr})
or (\ref{gferbr}) (with diagonal $g^{\nu\mu}(U)$) satisfies Tsarev 
"semi-Hamiltonian" property and so can be integrated by Tsarev's
method. Probably, all semi-Hamiltonian systems are in fact
Hamiltonian corresponding to some weakly nonlocal H.T.P.B.
with (maybe) an infinite number of terms in the  nonlocal tail.
Some investigation of this problem can be found in 
\cite{fer4}, \cite{bfer} but in general this problem is still 
open. Let us mention also that the examples of
non-diagonalizable Hamiltonian integrable (by inverse scattering methods)
systems (\ref{htsys}) were also investigated in 
\cite{fer5}-\cite{fer6}.

 As was pointed out in \cite{fer3}-\cite{fer4},
if the manifold ${\cal M}^{N}$ has a holonomic net of lines of
curvature the metric $g^{\nu\mu}(U)$ and all the operators
$w^{\nu}_{k\eta}$ can be written in the diagonal form in 
the corresponding coordinates $r^{\nu} = r^{\nu}(U)$.
Here we don't impose this requirement
and consider any brackets of Ferapontov type. 

 We will assume that the flows $w^{\nu}_{k\eta}(U) U^{\eta}_{X}$
in the nonlocal part of (\ref{gferbr}) are linearly independent
(with constant coefficients).
(The nonlocal part in (\ref{gferbr}) represents actually the 
non-degenerate quadratic form on the linear space generated by
$w^{\nu}_{k\eta}(U) U^{\eta}_{X}$, $k = 1,\dots, g$ written in the
canonical form with $e_{k} = \pm 1$.) As was pointed out by 
E.V.Ferapontov the local functional

$$H = \int h(U) dX$$
generates in this case the local flow with respect to the bracket
(\ref{gferbr}) if and only if 
the functional $H$ is a conservation law for any of the flows
$w^{\nu}_{k\eta}(U) U^{\eta}_{X}$ such that
the expressions

$$w^{\nu}_{k\eta}(U) U^{\eta}_{X} 
{\partial h \over \partial U^{\nu}}$$
represent the total derivatives with respect to $X$ of some functions 
$Q_{k}(U)$ for any $k$. 

 This fact is also true for more general weakly nonlocal Poisson
brackets having the form:

$$\{\varphi^{i}(x), \varphi_{j}(y)\} = \sum_{k=1}^{G} 
B^{ij}_{k}(\varphi, \varphi_{x}, \dots) \delta^{(k)}(x-y) + $$
\begin{equation}
\label{weaknon}
+ \sum_{k=1}^{g} e_{k} S^{i}_{k}(\varphi, \varphi_{x}, \dots)
\nu (x-y) S^{j}_{k}(\varphi, \varphi_{y}, \dots) 
\end{equation}
where $\delta^{(k)}(x-y) = (d/dx)^{k} \delta(x-y)$, $e_{k} = \pm 1$,
and the set $\{S^{i}_{k}(\varphi, \varphi_{x}, \dots)\}$ is linearly
independent.

 As far as we know the first example written precisely in this form
was the Sokolov bracket (\cite{sokolov})

$$\{\varphi(x), \varphi(y)\} = - \varphi_{x} \nu (x-y) \varphi_{y}$$
for the Krichever-Novikov equation

$$\varphi_{t} = \varphi_{xxx} - 
{3 \over 2} {\varphi_{xx}^{2} \over \varphi_{x}} + 
{h(\varphi) \over \varphi_{x}} $$
where 
$h(\varphi) = c_{3}\varphi^{3} + c_{2}\varphi^{2} + 
c_{1}\varphi + c_{0}$, with the Hamiltonian function

$$H = \int \left( {1 \over 2} {\varphi_{xx}^{2} \over \varphi_{x}^{2}}
+ {1 \over 3} {h(\varphi) \over \varphi_{x}^{2}} \right) dx $$

 As was established in \cite{malts99u}-\cite{malts99i} the flows
$S^{i}_{k}(\varphi, \varphi_{x},\dots)$ commute with each
other for any general bracket (\ref{weaknon}) and conserve the
corresponding Hamiltonian structure (\ref{weaknon}) on the phase 
space $\{\varphi^{i}(x)\}$ (this fact was 
important for the averaging procedure for such brackets
considered there). However, for general brackets
(\ref{weaknon}) they are not necessarily generated by the local 
Hamiltonian functions having the form

$$H = \int h (\varphi, \varphi_{x},\dots) dx $$

 Actually the brackets (\ref{weaknon}) are very common for so-called
"integrable systems" (like KdV or NLS) 
possessing the multi-Hamiltonian
structures connected by the recursion operator according to the
Lenard-Magri scheme (\cite{magri}). Such, it was proved in 
\cite{enrubor} that all the higher PB brackets for KdV given by the
recursion scheme starting from Gardner-Zakharov-Faddev bracket

$$\{\varphi(x), \varphi(y)\} = \delta^{\prime}(x-y) $$
and the Magri bracket

$$\{\varphi(x), \varphi(y)\} = - \delta^{\prime\prime\prime}(x-y) +
4 \varphi(x) \delta^{\prime}(x-y) + 2 \varphi_{x} \delta (x-y) $$
have exactly the form (\ref{weaknon}). 
In \cite{malnov00} the same fact was proved
for the case of NLS hierarchy and also the weakly nonlocal form
of the "negative" symplectic forms for KdV and NLS was established.

 The brackets (\ref{dnbrack}), (\ref{mferbr}) and (\ref{gferbr})
appear in these systems as the "dispersionless" limit of the
corresponding bracket (\ref{weaknon}) or, in more general case,
as a result of the averaging of (\ref{weaknon}) on the families
of quasiperiodic solutions of corresponding evolution system
connected with the Whitham method for slow modulations
of parameters 
(\cite{dubrnov1}-\cite{dubrnov3}, 
\cite{novmal93}-\cite{malts99i}). 

 We consider here the
compatible brackets of Ferapontov type and prove
the similar facts for the case of the non-degenerate pencils
(i.e. $det \, g^{\nu\mu}_{(0)} \neq 0$) with also some
non-degeneracy conditions for non-local part of 
${\hat J_{(0)}} + \lambda {\hat J}_{(1)}$. 

 Let us mention also that the wide classes of local pencils of
Hydrodynamic Type (DN-brackets) were investigated in many
details in \cite{dubrov1} (see also \cite{dubrov2},
\cite{dubrov3} and references therein) where they play the 
important role in the structure of Dubrovin-Frobenius
manifolds connected with solutions of WDVV equation for
Topological Field Theories. In the works \cite{fer4},
\cite{mokhov1} some important questions of weakly non-local 
pencils of Hydrodynamic Type (H.T.) were also considered.
In the papers \cite{mokhov2}, \cite{fercombr} 
also the generic diagonal
compatible flat pencils in terms of inverse scattering method
(see \cite{zakharov}, \cite{krichever}) were
discussed.

\section{On the canonical form and symplectic operator for the
general F-bracket.}
\setcounter{equation}{0}

 Let us formulate now the properties of the brackets (\ref{gferbr})
established in \cite{malnov00} which we will need in further 
consideration.

 Let us consider the bracket (\ref{gferbr}) with non-degenerate
tensor $g^{\nu\mu}(U)$. According to Ferapontov results we can 
represent it as a Dirac restriction of DN-bracket in $R^{N+g}$
to the submanifold ${\cal M}^{N} \subset R^{N+g}$ with flat normal 
connection. Let us fix the point $z \in {\cal M}^{N}$ and introduce
the corresponding loop space in the vicinity of $z$:

$$L({\cal M}^{N}, z) = \{\gamma : R^{1} \rightarrow {\cal M}^{N} :
\gamma (-\infty) = \gamma (+\infty) = z \in {\cal M}^{N}\}$$

 So we fix the boundary conditions for the functions $U^{\nu}(X)$
and require that these functions rapidly approach their boundary
values at $X \rightarrow \pm \infty$.

\vspace{0.5cm}

{\bf Theorem.} (A.Ya.Maltsev, S.P.Novikov)

{\it Let us consider the bracket (\ref{gferbr}) with non-degenerate
$g^{\nu\mu}(U)$ defined on the loop space $L({\cal M}^{N}, z)$. 
Consider the corresponding embedding ${\cal M}^{N} \subset R^{N+g}$
with flat normal connection. Consider the flat orthogonal coordinates
$Z^{I}$ in $R^{N+g}$ such that:

 1) $Z^{I}(z) = 0$, $I = 1,\dots,N+g$ and
the corresponding DN-bracket in $R^{N+g}$ has the form

$$\{Z^{I}(X), Z^{J}(Y)\} = E^{I} \delta^{IJ} \delta^{\prime}(X-Y) 
\,\, , \,\,\, I,J = 1, \dots, N+g \,\, , \,\,\, E^{I} = \pm 1 $$

 2) The first $N$ coordinates $Z^{1}, \dots, Z^{N}$ are tangential to
${\cal M}^{N}$ at the point $z$.

 3) The last $g$ coordinates $Z^{N+1}, \dots, Z^{N+g}$ are orthogonal
to ${\cal M}^{N}$ at the point $z$.

 Then:

 1) The bracket (\ref{gferbr}) has exactly $N$ local Casimirs
given by the functionals

$$N^{\nu} = \int_{-\infty}^{+\infty} n^{\nu}(X) dX $$
where $n^{\nu}(U)$ are the restrictions of the first $N$
(tangential to ${\cal M}^{N}$ at $z$) coordinates 
$Z^{1}, \dots, Z^{N}$ on ${\cal M}^{N}$.

 2) All the flows 

$$U^{\nu}_{t_{k}} = w^{\nu}_{k\eta}(U) U^{\eta}_{X} $$ 
are Hamiltonian with respect to (\ref{gferbr}) with local
Hamiltonian functionals 

$$H_{k} = \int_{-\infty}^{+\infty} h_{k}(X) dX \,\,\, ,
\,\,\, k = 1, \dots, g$$
where $h_{k}(U)$ are the restriction of the last $g$ coordinates
$Z^{N+k}$, $k = 1, \dots, g$ on ${\cal M}^{N}$ and 
$w^{\nu}_{k\eta}(U)$ are the Weingarten operators corresponding
to parallel vector fields ${\bf N}_{k}(U)$ in the normal bundle
with the normalization:

$${\bf N}_{k}(z) = (0, \dots, 0, -E^{N+k}, 0, \dots, 0)^{T} $$
(where $-E^{N+k}$ stays at the position $N+k$) in the coordinates
$Z^{1}, \dots, Z^{N+g}$ 
(so 
$\langle{\bf N}_{k}(U),{\bf N}_{l}(U)\rangle = E^{N+k}\delta_{kl}$
).\footnote{There is an arithmetic mistake in the journal variant
of \cite{malnov00} where the generated flows are written as
$E^{N+k} w^{\nu}_{k\eta}(U) U^{\eta}_{X}$.}

 3) The bracket (\ref{gferbr}) has in coordinates $n^{\nu}(U)$
the "Canonical form" corresponding to the space 
$L({\cal M}^{N}, z)$, i.e.

$$ \{n^{\nu}(X), n^{\mu}(Y)\} = \left( \epsilon^{\nu} \delta^{\nu\mu} 
-  \sum_{k=1}^{g} e_{k} f^{\nu}_{k}(n) f^{\mu}_{k}(n) \right)
\delta^{\prime}(X-Y) - $$

$$- \sum_{k=1}^{g} e_{k} \left( f^{\nu}_{k}(n) \right)_{X}
f^{\mu}_{k}(n) \delta (X-Y) +  \sum_{k=1}^{g} e_{k}
\left(f^{\nu}_{k}(n)\right)_{X} \nu (X-Y) 
\left( f^{\mu}_{k}(n) \right)_{Y} $$
where $\epsilon^{\nu} = E^{\nu}$, $\nu = 1, \dots, N$,
$e_{k} = E^{N+k}, k = 1, \dots, g$, $n^{\nu}(z) = 0$,
$f^{\nu}_{k}(z) = 0$ (i.e. $f^{\nu}_{k}(0) = 0$).
(Actually we have the equality 
$f^{\nu}_{k}(U) \equiv N^{\nu}_{k}(U)$, where $N^{\nu}_{k}(U)$
are the first $N$ components of the vectors ${\bf N}_{k}(U)$
in the coordinates $Z^{1}, \dots, Z^{N+g}$).  }

\vspace{0.5cm}

 Let us note here that the Casimirs and "Canonical functionals", 
as well as the Canonical forms depend on the phase space 
$L({\cal M}^{N}, z)$. So, if we don't fix the loop space it is
better to speak just about the $N+g$ canonical functions
(the restrictions of the flat coordinates from $R^{N+g}$)
playing the role of Casimirs or Canonical functionals depending
on the boundary conditions. (Also we will have many 
"Canonical forms" of the bracket (\ref{gferbr}) with different
$f^{\nu}_{k}(U)$ in this case).

 For the case of Mokhov-Ferapontov bracket the Canonical form
will be the (\ref{canmfer}) for any fixed space $L({\cal M}^{N}, z)$
(although the coordinates $n^{\nu}(U)$ will be different for 
different loop spaces).
The explicit form of canonical functional, which is the
momentum operator in this case, can be written then as

$$P = {1 \over c} \int_{-\infty}^{+\infty} \left(
1 - \sqrt{1 - c \sum_{\nu=1}^{N} \epsilon^{\nu} n^{\nu} n^{\nu}}
\right) dX $$
(\cite{malnov00}).

 Using the restriction of the symplectic form of DN-bracket on
${\cal M}^{N}$ it is also possible to get the symplectic form
$\Omega_{\nu\mu}(X,Y)$ for the F-bracket (\ref{gferbr}) with
non-degenerate $g^{\nu\mu}(U)$ (\cite{malnov00}). The symplectic
form appears to be also weakly nonlocal and can be written as

$$\Omega_{\nu\mu}(X,Y) = \sum_{I=1}^{N+g} E^{I}
{\partial Z^{I}(U) \over \partial U^{\nu}}(X) \nu (X-Y)
{\partial Z^{I}(U) \over \partial U^{\mu}}(Y) = $$

$$= \sum_{\tau=1}^{N} \epsilon^{\tau}
{\partial n^{\tau} \over \partial U^{\nu}}(X) \nu (X-Y)
{\partial n^{\tau} \over \partial U^{\mu}}(Y) +
\sum_{k=1}^{g} e_{k} {\partial h_{k} \over \partial U^{\nu}}(X)
\nu (X-Y) {\partial h_{k} \over \partial U^{\mu}}(Y) $$
on ${\cal M}^{N}$.

 We can write also the corresponding symplectic operator
${\hat \Omega}_{\nu\mu}$ on $L({\cal M}^{N}, z)$ as

\begin{equation}
\label{symomnm}
{\hat \Omega}_{\nu\mu} = \sum_{\tau=1}^{N} \epsilon^{\tau}
{\partial n^{\tau} \over \partial U^{\nu}} D^{-1}
{\partial n^{\tau} \over \partial U^{\mu}} +
\sum_{k=1}^{g} e_{k} {\partial h_{k} \over \partial U^{\nu}}
D^{-1} {\partial h_{k} \over \partial U^{\mu}} 
\end{equation}
with $D^{-1}$ defined in the skew-symmetric way.

 Let us introduce the functional space ${\cal V}_{0}(z)$ of the
vector-fields $\xi^{\nu}(X)$ on $L({\cal M}^{N}, z)$ rapidly
decreasing $\xi^{\nu}(X) \rightarrow 0$ at 
$X \rightarrow \pm \infty$. It is easy to see that all the
Hydrodynamic Type systems (\ref{htsys}) satisfy this
requirement as the vector fields on $L({\cal M}^{N}, z)$ 
and so belong to ${\cal V}_{0}(z)$. We can then naturally define 
the action of symplectic operator ${\hat \Omega}_{\nu\mu}$
on the space ${\cal V}_{0}(z)$.

 We will need also the functional space ${\cal F}_{0}(z)$ of
$1$-forms $\omega_{\nu}(X)$ on $L({\cal M}^{N}, z)$ such
that $\omega_{\nu}(X) \rightarrow 0$ (rapidly decreasing)
at $X \rightarrow \pm \infty$. Let us note here that  
${\hat \Omega}_{\nu\mu}\xi^{\mu}(X) \notin {\cal F}_{0}(z)$
in the general case.

 We will prove now by the direct calculation
that the symplectic form  ${\hat \Omega}_{\nu\mu}$
is the inverse of the Hamiltonian operator ${\hat J}_{F}^{\nu\mu}$
on the appropriate functional spaces.

\vspace{0.5cm}

{\bf Theorem 1.}

{\it 

 (I) We have on the functional space ${\cal V}_{0}(z)$ the
relation

$${\hat J}_{F}^{\nu\xi} {\hat \Omega}_{\xi\mu} = {\hat I}$$
where ${\hat I}$ is the identity operator on the space 
${\cal V}_{0}(z)$.

\vspace{0.3cm}

 (II) We have on the functional space ${\cal F}_{0}(z)$:

$${\hat \Omega}_{\nu\xi} {\hat J}_{F}^{\xi\mu} = {\hat I}$$
where ${\hat I}$ is the identity operator on ${\cal F}_{0}(z)$. }

\vspace{0.5cm}

Proof.

 (I) We have 

$${\hat J}_{F}^{\nu\xi} {\hat \Omega}_{\xi\mu} =
\left[ g^{\nu\xi}(U){d \over dX} + b^{\nu\xi}_{\eta}(U) U^{\eta}_{X}
+ \sum_{k=1}^{g} e_{k} w^{\nu}_{k\eta}(U) U^{\eta}_{X} D^{-1}
w^{\xi}_{k\zeta}(U) U^{\zeta}_{X}\right] \times $$

$$\times \left[ \sum_{\tau=1}^{N} \epsilon^{\tau}
{\partial n^{\tau} \over \partial U^{\xi} } D^{-1}
{\partial n^{\tau} \over \partial U^{\mu} } +
\sum_{l=1}^{g} e_{l} {\partial h^{l} \over \partial U^{\xi} }
D^{-1} {\partial h^{l} \over \partial U^{\mu} } \right] $$
where the operators $w^{\nu}_{k\eta}(U)$ correspond to the vector
fields ${\bf N}_{(k)}(U)$ such that 

\begin{equation}
\label{normcon}
{\bf N}_{(k)}(z) = (0,\dots,0,-e_{k},0,\dots,0)^{T}
\end{equation}
($-e_{k}$ stays at the position $N+k$) and we have identically
$\langle {\bf N}_{(k)}(U), {\bf N}_{(l)}(U) \rangle = 
e_{k} \delta_{kl}$.

 Since the integrals of $n^{\tau}(U)$ and $h^{l}(U)$
generate the local flows according to ${\hat J}_{F}^{\nu\mu}$
we have:

\begin{equation}
\label{qpx}
w^{\xi}_{k\zeta}(U) U^{\zeta}_{X}
{\partial n^{\tau} \over \partial U^{\xi}} \equiv
\left(q^{\tau}_{k}(U)\right)_{X} \,\, , \,\,\,\,\,
w^{\xi}_{k\zeta}(U) U^{\zeta}_{X}
{\partial h^{l} \over \partial U^{\xi}} \equiv
\left( p^{l}_{k}(U) \right)_{X}
\end{equation}
for some functions $q^{\tau}_{k}(U)$ and $p^{l}_{k}(U)$.
Moreover, let us consider $N$ tangential vectors to ${\cal M}_{N}$
in $R^{N+g}$ corresponding to the coordinates $U^{\nu}$:

\begin{equation}
\label{etang}
{\bf e}_{(\xi)}(U) = \left({\partial n^{1} \over \partial U^{\xi} },
\dots, {\partial n^{N} \over \partial U^{\xi} },
{\partial h^{1} \over \partial U^{\xi} }, \dots,
{\partial h^{g} \over \partial U^{\xi} } \right)^{T}
\end{equation}
and our parallel orthogonal vector fields ${\bf N}_{(k)}(U)$ defined 
by the condition (\ref{normcon})
in the coordinates $(Z^{1}, \dots, Z^{N+g})$;

 Since by the definition:

$${d \over dX} {\bf N}_{(k)}(U) = w^{\xi}_{k\zeta}U^{\zeta}_{X} 
{\bf e}_{(\xi)}(U)$$
we have from (\ref{qpx}) 

$$\left(q^{\tau}_{k}\right)_{X} = 
\left(N^{\tau}_{(k)}\right)_{X} \,\, , \,\,\,
\left(p^{l}_{k}\right)_{X} = \left(N^{N+l}_{(k)}\right)_{X} 
\,\, , \,\,\, \tau = 1,\dots,N \,\, , \,\,\, l = 1, \dots g $$
for the components of ${\bf N}_{(k)}(U)$.

 Let us normalize the functions $q^{\tau}_{k}(U)$ and
$p^{l}_{k}(U)$ such that

\begin{equation}
\label{norm}
q^{\tau}_{k}(z) = 0 \,\,\, , \,\,\,\,\, 
p^{l}_{k}(z) = 0 
\end{equation}

 We have then

\begin{equation}
\label{qpn}
q^{\tau}_{k}(U) = N^{\tau}_{(k)}(U) \,\,\, , \,\,\,\,\,
p^{l}_{k}(U) = N^{N+l}_{(k)}(U) + e_{k} \delta^{l}_{k}
\end{equation}

 Now using the equalities

$$\left(q^{\tau}_{k}\right)_{X} = {d \over dX} q^{\tau}_{k}
- q^{\tau}_{k} {d \over dX} \,\,\, , \,\,\,\,\,
\left(p^{l}_{k}\right)_{X} = {d \over dX} p^{l}_{k} - 
p^{l}_{k} {d \over dX}$$
on $L({\cal M}^{N},z)$ we can write

$${\hat J}_{F}^{\nu\xi} {\hat \Omega}_{\xi\mu}|_{{\cal V}_{0}(z)} 
= $$ 

$$= \sum_{\tau=1}^{N} \epsilon^{\tau} \left[ g^{\nu\xi} 
\left({\partial n^{\tau} \over \partial U^{\xi}}\right)_{X} +
b^{\nu\xi}_{\eta} U^{\eta}_{X} 
{\partial n^{\tau} \over \partial U^{\xi}} + \sum_{k=1}^{g}
e_{k} w^{\nu}_{k\eta} U^{\eta}_{X} D^{-1} {d \over dX} 
q^{\tau}_{k} \right] D^{-1} 
{\partial n^{\tau} \over \partial U^{\mu}} + $$

$$+ \sum_{l=1}^{g} e_{l} \left[ g^{\nu\xi} 
\left({\partial h^{l} \over \partial U^{\xi}}\right)_{X} +
b^{\nu\xi}_{\eta} U^{\eta}_{X}
{\partial h^{l} \over \partial U^{\xi}} + \sum_{k=1}^{g}
e_{k} w^{\nu}_{k\eta} U^{\eta}_{X} D^{-1} {d \over dX}
p^{l}_{k} \right] D^{-1}
{\partial h^{l} \over \partial U^{\mu}} + $$

$$+ g^{\nu\xi} \left[ \sum_{\tau=1}^{N} \epsilon^{\tau}
{\partial n^{\tau} \over \partial U^{\xi}} {d \over dX}
D^{-1} {\partial n^{\tau} \over \partial U^{\mu}} +
\sum_{l=1}^{g} e_{l} {\partial h^{l} \over \partial U^{\xi}}
{d \over dX} D^{-1}
{\partial h^{l} \over \partial U^{\mu}} \right] - $$

$$- \sum_{k=1}^{g} \sum_{\tau=1}^{N} e_{k} \epsilon^{\tau}
w^{\nu}_{k\eta} U^{\eta}_{X} D^{-1} q^{\tau}_{k} {d \over dX}
D^{-1} {\partial n^{\tau} \over \partial U^{\mu}} -
\sum_{k=1}^{g} \sum_{l=1}^{g} e_{k} e_{l} 
w^{\nu}_{k\eta} U^{\eta}_{X} D^{-1} p^{l}_{k} {d \over dX}
D^{-1} {\partial h^{l} \over \partial U^{\mu}} $$

 We can replace here $(d/dX) D^{-1}$ by identity and we have

$$\left( D^{-1} f_{X} \right) = f(X) - {1 \over 2} 
\left[ f(-\infty) + f(+\infty) \right] $$
for any function $f(U)$ on $L({\cal M}^{N},z)$
according to the definition of $D^{-1}$. So, according
to normalization (\ref{norm}) we can replace the operators
$D^{-1}(d/dX)q^{\tau}_{k}$ and $D^{-1}(d/dX)p^{l}_{k}$ just by
$q^{\tau}_{k}$ and $p^{l}_{k}$ on $L({\cal M}^{N},z)$.

 According to the same normalization the expressions 
within the brackets in the first two terms are equal to

\begin{equation}
\label{jwrel}
\left[ {\hat J}_{F}^{\nu\xi} 
{\partial n^{\tau} \over \partial U^{\xi}} 
\right]_{L({\cal M}^{N},z)} = 0 \,\,\, {\rm and} \,\,\,\,\,
\left[ {\hat J}_{F}^{\nu\xi}
{\partial h^{l} \over \partial U^{\xi}}
\right]_{L({\cal M}^{N},z)} = w^{\nu}_{l\eta} U^{\eta}_{X}
\end{equation}

 So we have

$${\hat J}_{F}^{\nu\xi} {\hat \Omega}_{\xi\mu}|_{{\cal V}_{0}(z)} 
= \sum_{l=1}^{g} e_{l} w^{\nu}_{l\eta} U^{\eta}_{X} D^{-1}
{\partial h^{l} \over \partial U^{\mu}} + 
g^{\nu\xi} \left[ \sum_{\tau=1}^{N} \epsilon^{\tau}
{\partial n^{\tau} \over \partial U^{\xi}}
{\partial n^{\tau} \over \partial U^{\mu}} +
\sum_{l=1}^{g} e_{l} {\partial h^{l} \over \partial U^{\xi}}
{\partial h^{l} \over \partial U^{\mu}} \right] - $$

$$- \sum_{k=1}^{g} e_{k} w^{\nu}_{k\eta} U^{\eta}_{X} D^{-1}
\left[ \sum_{\tau=1}^{N} \epsilon^{\tau} q^{\tau}_{k}
{\partial n^{\tau} \over \partial U^{\mu}} +
\sum_{l=1}^{g} e_{l} p^{l}_{k}
{\partial h^{l} \over \partial U^{\mu}} \right] $$

 Using (\ref{qpn}) and (\ref{etang}) we get now

\begin{equation}
\label{qprel}
\sum_{\tau=1}^{N} \epsilon^{\tau} q^{\tau}_{k}
{\partial n^{\tau} \over \partial U^{\mu}} +
\sum_{l=1}^{g} e_{l} p^{l}_{k}
{\partial h^{l} \over \partial U^{\mu}} = 
\langle {\bf N}_{(k)}, {\bf e}_{(\mu)} \rangle + 
{\partial h^{k} \over \partial U^{\mu}} = 
{\partial h^{k} \over \partial U^{\mu}} 
\end{equation}

 Using also the evident relation

$$\sum_{\tau=1}^{N} \epsilon^{\tau}
{\partial n^{\tau} \over \partial U^{\xi}}
{\partial n^{\tau} \over \partial U^{\mu}} +
\sum_{l=1}^{g} e_{l} {\partial h^{l} \over \partial U^{\xi}}
{\partial h^{l} \over \partial U^{\mu}} \equiv g_{\xi\mu}(U)$$
we get the statement (I) of the theorem.

\vspace{0.5cm}

 (II) We have

$${\hat \Omega}_{\nu\xi} {\hat J}_{F}^{\xi\mu} = 
\left( \sum_{\tau=1}^{N} \epsilon^{\tau}
{\partial n^{\tau} \over \partial U^{\nu}} D^{-1} 
{\partial n^{\tau} \over \partial U^{\xi}} +
\sum_{l=1}^{g} e_{l} {\partial h^{l} \over \partial U^{\nu}}
D^{-1} {\partial h^{l} \over \partial U^{\xi}} \right) \times $$

$${\hfill \times \left( g^{\xi\mu} {d \over dX} +
b^{\xi\mu}_{\eta} U^{\eta}_{X} + \sum_{k=1}^{g} e_{k}
w^{\xi}_{k\eta} U^{\eta}_{X} D^{-1} 
w^{\mu}_{k\zeta} U^{\zeta}_{X} \right) = }$$

$$= \sum_{\tau=1}^{N} \epsilon^{\tau}
{\partial n^{\tau} \over \partial U^{\nu}} D^{-1}
{d \over dX} {\partial n^{\tau} \over \partial U^{\xi}}
g^{\xi\mu} +  \sum_{l=1}^{g}
{\partial h^{l} \over \partial U^{\nu}} D^{-1}
{d \over dX} {\partial h^{l} \over \partial U^{\xi}}
g^{\xi\mu} - $$

$$- \sum_{\tau=1}^{N} \epsilon^{\tau}
{\partial n^{\tau} \over \partial U^{\nu}} D^{-1} \left( 
{\partial n^{\tau} \over \partial U^{\xi}} g^{\xi\mu} \right)_{X}
- \sum_{l=1}^{g}
{\partial h^{l} \over \partial U^{\nu}} D^{-1} \left(
{\partial h^{l} \over \partial U^{\xi}} g^{\xi\mu} \right)_{X} +$$

$$+ \sum_{\tau=1}^{N} \epsilon^{\tau}
{\partial n^{\tau} \over \partial U^{\nu}} D^{-1}
{\partial n^{\tau} \over \partial U^{\xi}} 
b^{\xi\mu}_{\eta} U^{\eta}_{X} + \sum_{l=1}^{g} e_{l} 
{\partial h^{l} \over \partial U^{\nu}} D^{-1}
{\partial h^{l} \over \partial U^{\xi}} 
b^{\xi\mu}_{\eta} U^{\eta}_{X} + $$

$$+ \sum_{\tau=1}^{N} \sum_{k=1}^{g} \epsilon^{\tau} e_{k}
{\partial n^{\tau} \over \partial U^{\nu}} D^{-1}
\left( {d \over dX} q^{\tau}_{k} - q^{\tau}_{k} {d \over dX} 
\right) D^{-1} w^{\mu}_{k\zeta} U^{\zeta}_{X} + $$

$$+ \sum_{l=1}^{g} \sum_{k=1}^{g} e_{l} e_{k}
{\partial h^{l} \over \partial U^{\nu}} D^{-1}
\left( {d \over dX} p^{l}_{k} - p^{l}_{k} {d \over dX} \right)
D^{-1} w^{\mu}_{k\zeta} U^{\zeta}_{X} $$

 We can replace again the operators $(d/dX) D^{-1}$ by identity
and \linebreak
$D^{-1}(d/dX)q^{\tau}_{k}$ and $D^{-1}(d/dX)p^{l}_{k}$ by
$q^{\tau}_{k}$ and $p^{l}_{k}$. Then according to the definition
of coordinates $h^{l}$ we have 
$(\partial h^{l}/\partial U^{\xi})(z) = 0$, so we put also
$D^{-1}(d/dX)(\partial h^{l}/\partial U^{\xi}) =
(\partial h^{l}/\partial U^{\xi})$. We get now using the same
arguments

$${\hat \Omega}_{\nu\xi} {\hat J}_{F}^{\xi\mu} = {\hat I} +
\sum_{\tau=1}^{N} \epsilon^{\tau} 
{\partial n^{\tau} \over \partial U^{\nu}} 
\left[  D^{-1} {d \over dX} - {\hat I} \right] 
{\partial n^{\tau} \over \partial U^{\xi}}
g^{\xi\mu} - $$

$$- \sum_{\tau=1}^{N} \epsilon^{\tau}
{\partial n^{\tau} \over \partial U^{\nu}} D^{-1}
\left[ {\hat J}_{F}^{\mu\xi}
{\partial n^{\tau} \over \partial U^{\xi}} 
\right]_{L({\cal M}^{N},z)} - \sum_{l=1}^{g} e_{l}
{\partial h^{l} \over \partial U^{\nu}} D^{-1}
\left[ {\hat J}_{F}^{\mu\xi}
{\partial h^{l} \over \partial U^{\xi}} 
\right]_{L({\cal M}^{N},z)} + $$

$$+ \sum_{k=1}^{g} \left( \sum_{\tau=1}^{N} \epsilon^{\tau}
{\partial n^{\tau} \over \partial U^{\nu}} q^{\tau}_{k} D^{-1} 
w^{\mu}_{k\zeta} U^{\zeta}_{X} + \sum_{l=1}^{g} e_{l}
{\partial h^{l} \over \partial U^{\nu}} p^{l}_{k} D^{-1}
w^{\mu}_{k\zeta} U^{\zeta}_{X} \right) $$
(we used the skew-symmetry of the operator 
${\hat J}_{F}^{\mu\xi}$ for the action from the right).

 Using again (\ref{jwrel}) and (\ref{qprel}) we get

$${\hat \Omega}_{\nu\xi} {\hat J}_{F}^{\xi\mu} = {\hat I} +
\sum_{\tau=1}^{N} \epsilon^{\tau}
{\partial n^{\tau} \over \partial U^{\nu}}
\left[  D^{-1} {d \over dX} - {\hat I} \right] 
{\partial n^{\tau} \over \partial U^{\xi}} g^{\xi\mu} $$
i.e

$${\hat \Omega}_{\nu\xi} {\hat J}_{F}^{\xi\mu} f_{\mu}(X) =
f_{\mu}(X) - \sum_{\tau=1}^{N} \epsilon^{\tau}
{\partial n^{\tau} \over \partial U^{\nu}}(X)
{\partial n^{\tau} \over \partial U^{\xi}}(z) g^{\xi\mu}(z)
f_{\mu}(z) $$
for any $f_{\mu}(X)$. From the definition of ${\cal F}_{0}(z)$
we obtain now the part (II) of the theorem.

{\hfill Theorem is proved.} 

\vspace{0.5cm}

 {\bf Remark.} We can say also that the operator 
${\hat J}_{F}^{\nu\xi} {\hat \Omega}_{\xi\mu}$ is identity
if acting from the left on ${\cal F}_{0}(z)$ and from
the right on ${\cal V}_{0}(z)$. This interpretation will
be convenient later for the consideration of the recursion
operator.

\vspace{0.5cm}

 Let us introduce also the Momentum functional $P$ generating
the flow 

\begin{equation}
\label{shift}
U^{\nu}_{T} = U^{\nu}_{X}
\end{equation}
with respect to the general bracket (\ref{gferbr}) 
(with non-degenerate $g^{\nu\mu}(U)$).

\vspace{0.5cm}

{\bf Lemma 1.}

{\it Any F-bracket (\ref{gferbr}) with non-degenerate tensor
$g^{\nu\mu}(U)$ has the local Momentum operator $P$ generating
the flow (\ref{shift}) on the space $L({\cal M}^{N},z)$. The 
functional $P$ can be written in the form

\begin{equation}
\label{momfer}
P = \int_{-\infty}^{+\infty} p(U) dX =
{1 \over 2} \int_{-\infty}^{+\infty} \left(
\sum_{\tau=1}^{N} \epsilon^{\tau} n^{\tau} n^{\tau} +
\sum_{k=1}^{g} e_{k} h^{k} h^{k} \right) dX
\end{equation}
where the functions $n^{\tau}$ and $h^{k}$ correspond to the loop
space $L({\cal M}^{N},z)$. }

\vspace{0.5cm}

 Proof.

We should just prove here the relation

$${\partial p \over \partial U^{\nu}} = {\hat \Omega}_{\nu\xi}
U^{\xi}_{X} $$ 
on $L({\cal M}^{N},z)$ according to part (I) of Theorem 1. 
So we have

$$\sum_{\tau=1}^{N} \epsilon^{\tau}
{\partial n^{\tau} \over \partial U^{\nu}} D^{-1}
{\partial n^{\tau} \over \partial U^{\xi}} U^{\xi}_{X} +
\sum_{k=1}^{g} e_{k}
{\partial h^{k} \over \partial U^{\nu}} D^{-1}
{\partial h^{k} \over \partial U^{\xi}} U^{\xi}_{X} = $$

$$= \sum_{\tau=1}^{N} \epsilon^{\tau}
{\partial n^{\tau} \over \partial U^{\nu}} n^{\tau} +
\sum_{k=1}^{g} e_{k}
{\partial h^{k} \over \partial U^{\nu}} h^{k} = 
{\partial \over \partial U^{\nu}} p $$

{\hfill Lemma is proved.}

\section{The $\lambda$-pencils and the integrable hierarchies.}
\setcounter{equation}{0}

 Let us consider now the operator 

$${\hat J}_{\lambda}^{\nu\mu} = {\hat J}_{(0)}^{\nu\mu} +
\lambda {\hat J}_{(1)}^{\nu\mu} = 
\left(g^{\nu\mu}_{(0)} + \lambda g^{\nu\mu}_{(1)} \right)
{d \over dX} + \left( b^{\nu\mu}_{(0)\eta} +
\lambda b^{\nu\mu}_{(1)\eta} \right) U^{\eta}_{X} + $$
\begin{equation}
\label{lpencil}
+ \sum_{k=1}^{g_{0}} e_{(0)k} w^{\nu}_{(0)k\eta} U^{\eta}_{X}
D^{-1} w^{\mu}_{(0)k\zeta} U^{\zeta}_{X} + \lambda
\sum_{k=1}^{g_{1}} e_{(1)k} w^{\nu}_{(1)k\eta} U^{\eta}_{X}
D^{-1} w^{\mu}_{(1)k\zeta} U^{\zeta}_{X}
\end{equation}
for ${\hat J}_{(0)}^{\nu\mu}$ and ${\hat J}_{(1)}^{\nu\mu}$
having the form (\ref{gferop}). We will call
the $\lambda$-pencil (\ref{lpencil}) non-degenerate
(for small $\lambda$) if $det \, g^{\nu\mu}_{(0)}(U) \neq 0$.

 We will admit here that the linear spaces ${\cal W}_{(0)}$
and ${\cal W}_{(1)}$ generated by the sets

$$\{w^{\nu}_{(0)1\eta} U^{\eta}_{X}, \dots, 
w^{\nu}_{(0)g_{0}\eta} U^{\eta}_{X}\} \,\,\, {\rm and}
\,\,\, \{w^{\nu}_{(1)1\eta} U^{\eta}_{X}, \dots,
w^{\nu}_{(1)g_{1}\eta} U^{\eta}_{X}\} $$
can have a nontrivial intersection ${\cal V}$.

 Let us introduce the basis in ${\cal V}$

$$\{ {\hat v}^{\nu}_{1\eta} U^{\eta}_{X}, \dots ,
{\hat v}^{\nu}_{d\eta} U^{\eta}_{X} \} $$
where $d = dim \, {\cal V}$ and consider the linear space
${\cal W}$ generated by all the flows from ${\cal W}_{(0)}$
and ${\cal W}_{(1)}$ ( $dim \, {\cal W} = g_{0} + g_{1} - d$)
with basis 

$${\cal B} = \{{\hat w}^{\nu}_{(0)1\eta} U^{\eta}_{X}, \dots,
{\hat w}^{\nu}_{(0)(g_{0}-d)\eta} U^{\eta}_{X},
{\hat v}^{\nu}_{1\eta} U^{\eta}_{X}, \dots ,
{\hat v}^{\nu}_{d\eta} U^{\eta}_{X},
{\hat w}^{\nu}_{(1)1\eta} U^{\eta}_{X}, \dots,
{\hat w}^{\nu}_{(1)(g_{1}-d)\eta} U^{\eta}_{X} \} = $$

\begin{equation}
\label{wbasis}
= \{{\tilde w}^{\nu}_{1\eta}(U) U^{\eta}_{X}, \dots ,
{\tilde w}^{\nu}_{g_{0}+g_{1}-d,\eta}(U) U^{\eta}_{X} \}
\end{equation}
where ${\hat w}^{\nu}_{(0)k\eta}$ and 
${\hat w}^{\nu}_{(1)s\eta}$ are some linear combinations of
operators $w_{(0)}$ and $w_{(1)}$ respectively, and the
flows corresponding to
${\hat w}_{(0)k}$, ${\hat v}_{m}$ and ${\hat w}_{(1)s}$
are all linearly independent.

 The flows

$$\{{\hat w}^{\nu}_{(0)1\eta} U^{\eta}_{X}, \dots,
{\hat w}^{\nu}_{(0)(g_{0}-d)\eta} U^{\eta}_{X},
{\hat v}^{\nu}_{1\eta} U^{\eta}_{X}, \dots ,
{\hat v}^{\nu}_{d\eta} U^{\eta}_{X}\} $$
and

$$\{{\hat v}^{\nu}_{1\eta} U^{\eta}_{X}, \dots ,
{\hat v}^{\nu}_{d\eta} U^{\eta}_{X},
{\hat w}^{\nu}_{(1)1\eta} U^{\eta}_{X}, \dots,
{\hat w}^{\nu}_{(1)(g_{1}-d)\eta} U^{\eta}_{X} \} $$
will give then bases in ${\cal W}_{(0)}$ and ${\cal W}_{(1)}$
respectively.

 The nonlocal part of the bracket ${\hat J}^{\nu\mu}_{\lambda}$

$$\sum_{k=1}^{g_{0}} e_{(0)k} w^{\nu}_{(0)k\eta} U^{\eta}_{X}
D^{-1} w^{\mu}_{(0)k\zeta} U^{\zeta}_{X} + \lambda
\sum_{s=1}^{g_{1}} e_{(1)s} w^{\nu}_{(1)s\eta} U^{\eta}_{X}
D^{-1} w^{\mu}_{(1)s\zeta} U^{\zeta}_{X} $$
will correspond in our case to some quadratic form
$Q_{\lambda}^{ks}$ (linear in $\lambda$), 
$k,s = 1,\dots,g_{0}+g_{1}-d$ on the space ${\cal W}$.

 In our further consideration the question if $Q_{\lambda}^{ks}$
is non-degenerate on ${\cal W}$ for $\lambda \neq 0$ or not will 
be important and
we will mainly consider the pencils (\ref{lpencil}) such that
$Q_{\lambda}^{ks}$ is non-degenerate on ${\cal W}$.

 Let us formulate now the properties of non-degenerate
pencils (\ref{lpencil}) satisfying also the requirement of 
non-degeneracy of form $Q_{\lambda}^{ks}$ for $\lambda \neq 0$
connected with the "canonical" integrable hierarchies.

\vspace{0.5cm}

{\bf Theorem 2.}

{\it Let us consider the non-degenerate pencil (\ref{lpencil})
($det \, g^{\nu\mu}_{(0)} \neq 0$) such that the form 
$Q_{\lambda}^{ks}$ is also non-degenerate on ${\cal W}$ for 
$\lambda \neq 0$ (small enough). Then:

\vspace{0.3cm}

(I) It is possible to introduce the local functionals:

\begin{equation}
\label{anpen}
N^{\nu}(\lambda) = \int_{-\infty}^{+\infty} 
n^{\nu}(U,\lambda) dX \,\,\, , \,\,\,\,\, \nu = 1, \dots, N
\end{equation}

\begin{equation}
\label{mompen}
P (\lambda) = \int_{-\infty}^{+\infty}
p (U,\lambda) dX 
\end{equation}

\begin{equation}
\label{h0pen}
H^{k}_{(0)} (\lambda) = \int_{-\infty}^{+\infty}
h^{k}_{(0)} (U,\lambda) dX \,\,\, , \,\,\,\,\, k = 1, \dots, g_{0}    
\end{equation}
and 

\begin{equation}
\label{h1pen}
H^{s}_{(1)} (\lambda) = \int_{-\infty}^{+\infty}
h^{s}_{(1)} (U,\lambda) dX \,\,\, , \,\,\,\,\, s = 1, \dots, g_{1}
\end{equation}
which are the Casimirs, Momentum operator and the Hamiltonian
functions for the flows $w^{\nu}_{(0)k\eta}(U) U^{\eta}_{X}$
and $w^{\nu}_{(1)s\eta}(U) U^{\eta}_{X}$ for the bracket
${\hat J}^{\nu\mu}_{\lambda}$ respectively.

\vspace{0.3cm}

 (II) All the functions $n^{\nu}(U,\lambda)$, $p(U,\lambda)$,
$h^{k}_{(0)}(U,\lambda)$ and $h^{s}_{(1)}(U,\lambda)$ are regular 
at $\lambda \rightarrow 0$ and can be represented as regular 
series:

\begin{equation}
\label{expnp}
n^{\nu}(U,\lambda) = \sum_{q=0}^{+\infty} n^{\nu}_{q}(U)
\lambda^{q} \,\,\, , \,\,\,\,\,
p(U,\lambda) = \sum_{q=0}^{+\infty} p_{q}(U) \lambda^{q} 
\end{equation}

\begin{equation}
\label{exphh}
h^{k}_{(0)}(U,\lambda) = \sum_{q=0}^{+\infty}
h^{k}_{(0),q}(U) \lambda^{q} \,\,\, , \,\,\,\,\,
h^{s}_{(1)}(U,\lambda) = \sum_{q=0}^{+\infty}
h^{s}_{(1),q}(U) \lambda^{q}
\end{equation}

 Moreover, we can choose these functionals in such a way that:

$${\partial p_{q} \over \partial U^{\mu}}(z) = 0 \,\, , \,\,\,
{\partial h^{k}_{(0),q} \over \partial U^{\mu}}(z) = 0 \,\, , \,\,\,
{\partial h^{s}_{(1),q} \over \partial U^{\mu}}(z) = 0 \,\, , \,\,\,
q \geq 0 $$

$${\partial n^{\nu}_{q} \over \partial U^{\mu}}(z) = 0 \,\, , \,\,\,
q \geq 1 \,\,\, {\rm and} \,\,\, 
{\partial n^{\nu}_{0} \over \partial U^{\mu}}(z) = e^{\nu}_{\mu} $$
where $det \, e^{\nu}_{\mu} \neq 0$.

\vspace{0.3cm}

 (III) The integrals $N^{\nu}(0)$, $P(0)$, $H^{k}_{(0)}(0)$ and
$H^{s}_{(1)}(0)$ are the Casimirs, Momentum operator and the 
Hamiltonian functions for the flows 
$w^{\nu}_{(0)k\eta}(U) U^{\eta}_{X}$ and
$w^{\nu}_{(1)s\eta}(U) U^{\eta}_{X}$ with respect to 
${\hat J}_{(0)}^{\nu\mu}$, while the flows generated
by the functionals

$$F_{q} = \int_{-\infty}^{+\infty} f_{q}(U) dX $$
are connected by the relation:

$${\hat J}_{(0)}^{\nu\xi} 
{\partial f_{(q+1)} \over \partial U^{\xi}} = -
{\hat J}_{(1)}^{\nu\xi} 
{\partial f_{q} \over \partial U^{\xi}} $$
for any functional $F(\lambda)$ from the set 
(\ref{anpen})-(\ref{h1pen}). All the functionals $F_{q}$
given by the expansions (\ref{expnp})-(\ref{exphh}) generate
the local flows and commute with each other with respect to
both brackets ${\hat J}_{(0)}$ and ${\hat J}_{(1)}$.
}

\vspace{0.5cm}

 Proof.

 Let us put now $\lambda > 0$ and write 
${\hat J}_{\lambda}^{\nu\mu}$ in the form:

$${\hat J}_{\lambda}^{\nu\mu} =
\left(g^{\nu\mu}_{(0)} + \lambda g^{\nu\mu}_{(1)} \right)
{d \over dX} + \left( b^{\nu\mu}_{(0)\eta} +
\lambda b^{\nu\mu}_{(1)\eta} \right) U^{\eta}_{X} + $$

\begin{equation}
\label{diagjnml}
+ \sum_{k=1}^{g_{0}} e_{(0)k} w^{\nu}_{(0)k\eta} U^{\eta}_{X}
D^{-1} w^{\mu}_{(0)k\zeta} U^{\zeta}_{X} + \sum_{k=1}^{g_{1}}
e_{(1)k} \sqrt{\lambda} w^{\nu}_{(1)k\eta} U^{\eta}_{X}
D^{-1} \sqrt{\lambda} w^{\mu}_{(1)k\zeta} U^{\zeta}_{X}
\end{equation}

 In the case of the non-degenerate form $Q_{\lambda}^{ks}$
on ${\cal W}$ we can consider the corresponding embedding of
${\cal M}^{N}$ to $R^{N+g_{0}+g_{1}}$ depending on $\lambda$.
Indeed, all the flows ${\tilde w}_{s}$ from the set ${\cal B}$
will satisfy in this case to conditions

$$g^{\nu\xi}_{\lambda} {\tilde w}^{\mu}_{s\xi} =
g^{\mu\xi}_{\lambda} {\tilde w}^{\nu}_{s\xi} \,\,\, , \,\,\,
\nabla_{\nu} {\tilde w}^{\mu}_{s\eta} = 
\nabla_{\eta} {\tilde w}^{\mu}_{s\nu} $$

$$R^{\nu\mu}_{\eta\zeta} = \sum_{k,s} \left(
{\tilde w}^{\nu}_{k\eta} Q_{\lambda}^{ks} 
{\tilde w}^{\mu}_{s\zeta} - 
{\tilde w}^{\mu}_{k\eta} Q_{\lambda}^{ks}
{\tilde w}^{\nu}_{s\zeta} \right)$$
for non-degenerate $g^{\mu\xi}_{\lambda}$ according to
Ferapontov theorem.

 Since the operators $w^{\nu}_{(0)k\eta}(U)$,
$w^{\nu}_{(1)s\eta}(U)$ are just linear combinations of
${\tilde w}^{\nu}_{n\eta}(U)$ (with constant coefficients)
and the form $Q_{\lambda}^{ks}$ coincides with the nonlocal part of 
(\ref{diagjnml}) we will have the corresponding Gauss and
Petersson-Codazzi equations for these flows and the curvature tensor 
$R^{\nu\mu}_{\eta\zeta}$. So for non-degenerate 
$g^{\nu\mu}_{\lambda}$ we will get the local embedding of
${\cal M}^{N}$ to $R^{N+g_{0}+g_{1}}$ (actually to some subspace
$R^{N+g_{0}+g_{1}-d} \subset R^{N+g_{0}+g_{1}}$) depending
on $\lambda$.

 Since the embedding is defined up to the Poincare transformation
in $R^{N+g_{0}+g_{1}}$ we can choose at all $\lambda$ the 
coordinates $Z^{I}$, $I = 1, \dots, N+g_{0}+g_{1}$ in such a way
that:

 1) All $Z^{I} = 0$ at the point $z \in {\cal M}^{N}$ and the
metric $G^{IJ}$ in $R^{N+g_{0}+g_{1}}$ has the form

$$G^{IJ} = E^{I} \delta^{IJ}$$
where $E^{N+k} = e_{(0)k}$, $k = 1,\dots,g_{0}$, 
$E^{N+g_{0}+s} = e_{(1)s}$, $s = 1,\dots,g_{1}$.

 2) The first $N$ coordinates $Z^{\nu}$, $\nu = 1,\dots,N$
are tangential to ${\cal M}^{N}$ at the point $z \in {\cal M}^{N}$
at all $\lambda$.

 3) The last $g_{0}+g_{1}$ coordinates are orthogonal to
${\cal M}^{N}$ at the point $z$ and the Weingarten operators

$$w^{\nu}_{(0)1\eta}(U) \, , \dots \, , \,
w^{\nu}_{(0)g_{0}\eta}(U) \, , 
\sqrt{\lambda} w^{\nu}_{(1)1\eta}(U)
\, , \dots \, , \, \sqrt{\lambda} w^{\nu}_{(1)g_{1}\eta}(U)$$
correspond to the parallel vector fields ${\bf N}_{(0)(k)}(U)$,
${\bf N}_{(1)(s)}(U)$ in the normal bundle such that:

$${\bf N}_{(0)(k)}(z) = 
\left(0, \dots, 0, - E^{N+k}, 0,\dots, 0 \right)^{T} $$
($- E^{N+k}$ stays at the position $N+k$) $k = 1, \dots, g_{0}$, 

$${\bf N}_{(1)(k)}(z) =
\left(0, \dots, 0, - E^{N+g_{0}+s}, 0,\dots, 0 \right)^{T} $$
($- E^{N+g_{0}+s}$ stays at the position $N+g_{0}+s$)
$s = 1, \dots, g_{1}$.

 So, according to \cite{malnov00} the restriction of the first 
$N$ coordinates $Z^{1}, \dots, Z^{N}$ gives the Casimirs of
the bracket (\ref{diagjnml}) on $L({\cal M}^{N},z)$

\begin{equation}
\label{annlnu}
{\tilde N}^{\nu}(\lambda) = \int_{-\infty}^{+\infty}
{\tilde n}^{\nu}(U,\lambda) dX = \int_{-\infty}^{+\infty}
Z^{\nu}|_{{\cal M}^{N}}(U,\lambda) dX \,\,\, , \,\,\,
\nu = 1, \dots, N
\end{equation}
while the restrictions of the last $g_{0}+g_{1}$ coordinates 
give the Hamiltonian functions for the flows 
$w^{\nu}_{(0)k\eta} U^{\eta}_{X}$

\begin{equation}
\label{h0klnu}
{\tilde H}^{k}_{(0)}(\lambda) = \int_{-\infty}^{+\infty}
{\tilde h}^{k}_{(0)}(U,\lambda) dX = \int_{-\infty}^{+\infty}
Z^{N+k}|_{{\cal M}^{N}}(U,\lambda) dX \,\,\, , \,\,\,  
k = 1, \dots, g_{0} 
\end{equation}
and $\sqrt{\lambda} w^{\nu}_{(1)s\eta} U^{\eta}_{X}$ 

\begin{equation}
\label{h1slnu}
{\tilde H}^{s}_{(1)}(\lambda) = \int_{-\infty}^{+\infty}
{\tilde h}^{s}_{(1)}(U,\lambda) dX = \int_{-\infty}^{+\infty}
Z^{N+g_{0}+s}|_{{\cal M}^{N}}(U,\lambda) dX \,\,\, , \,\,\,
s = 1, \dots, g_{1}
\end{equation}

 Remark. For $\lambda < 0$ the signature of $R^{N+g_{0}+g_{1}}$
may be different from the case $\lambda > 0$ but all the statements
of the Theorem will certainly be also true.

 Let us study now the $\lambda$-dependence of the functions
${\tilde n}^{\nu}(U,\lambda)$, ${\tilde h}^{k}_{(0)}(U,\lambda)$
and ${\tilde h}^{s}_{(1)}(U,\lambda)$ at $\lambda \rightarrow 0$.

 Consider $N$ tangential vectors to ${\cal M}^{N}$ corresponding 
to the coordinate system $\{U^{\nu}\}$, i.e.

$${\bf e}_{(\nu)} = \left({\partial Z^{1} \over \partial U^{\nu} },
\dots, {\partial Z^{N+g_{0}+g_{1}} \over \partial U^{\nu} }
\right)^{T} \,\,\, , \,\,\, \nu = 1, \dots, N $$

 We have the following relations (in $R^{N+g_{0}+g_{1}}$)
for the differentials of
${\bf e}_{(\nu)}(U,\lambda)$, ${\bf N}_{(0)(k)}(U,\lambda)$
and ${\bf N}_{(1)(s)}(U,\lambda)$ on ${\cal M}^{N}$:

$$d {\bf e}_{(\nu)} = \Gamma^{\mu}_{\nu\eta}(U,\lambda)
{\bf e}_{(\mu)} dU^{\eta} - $$

$$- \sum_{k=1}^{g_{0}} E^{N+k} g_{\nu\mu}(U,\lambda)
w^{\mu}_{(0)k\eta}(U,\lambda) {\bf N}_{(0)(k)} dU^{\eta} -
\sqrt{\lambda} \sum_{s=1}^{g_{1}} E^{N+g_{0}+s}
g_{\nu\mu}(U,\lambda) w^{\mu}_{(1)s\eta}(U,\lambda)
{\bf N}_{(1)(s)} dU^{\eta} $$

$$d {\bf N}_{(0)(k)} = w^{\nu}_{(0)k\eta}(U,\lambda)
{\bf e}_{(\nu)} dU^{\eta} $$

$$d {\bf N}_{(1)(s)} = \sqrt{\lambda}
w^{\nu}_{(1)s\eta}(U,\lambda) {\bf e}_{(\nu)} dU^{\eta} $$
where $\Gamma^{\mu}_{\nu\eta} = -g_{\nu\xi} b^{\xi\mu}_{\eta}$,
$g_{\nu\xi} g^{\xi\mu} \equiv \delta^{\mu}_{\nu}$.

 So for any curve $\gamma(t)$ on ${\cal M}^{N}$ $(\gamma(0)=z)$
we have the evolution system for ${\bf e}_{(\nu)}(t)$, 
${\bf N}_{(0)(k)}(t)$ and ${\bf N}_{(1)(s)}(t)$ having the
general form:

\begin{equation}
\label{zvezds}
{d \over dt} \left( \begin{array}{c}
{\bf e}_{(\nu)}(t) \cr 
{\bf N}_{(0)(k)}(t) \cr
{\bf N}_{(1)(s)}(t) \end{array} \right) =
\left( \begin{array}{ccc}
* (t,\lambda) & * (t,\lambda) & \sqrt{\lambda} * (t,\lambda) \cr
* (t,\lambda) & 0 & 0 \cr
\sqrt{\lambda} * (t,\lambda) & 0 & 0 \end{array} \right)
\left( \begin{array}{c}
{\bf e}_{(\nu)}(t) \cr
{\bf N}_{(0)(k)}(t) \cr
{\bf N}_{(1)(s)}(t) \end{array} \right)
\end{equation}
where all $* (t,\lambda)$ are regular at $\lambda \rightarrow 0$
(matrix-) functions of $\lambda$.
 
 The formal solution of (\ref{zvezds}) can be written as the 
chronological exponent

$$T \exp \left( \int_{0}^{t} {\hat H}(t) dt \right) =
{\hat I} + \sum_{n=1}^{+\infty} {1 \over n!}
\int_{0}^{t_{1}} \dots \int_{0}^{t_{n}} T \left(
{\hat H}(t_{1}) \dots {\hat H}(t_{n}) \right)
dt_{1} \dots dt_{n}$$ 
applied to the initial data ${\bf e}_{(\nu)}(0)$, 
${\bf N}_{(0)(k)}(0)$, ${\bf N}_{(1)(s)}(0)$, where 
${\hat H}(t)$ is the matrix of the system (\ref{zvezds}).

 It is easy to verify now that for any $n \geq 1$ we have

$$T \left( {\hat H}(t_{1}) \dots {\hat H}(t_{n}) \right) =
\left( \begin{array}{ccc}  
* (t,\lambda) & * (t,\lambda) & \sqrt{\lambda} * (t,\lambda) \cr
* (t,\lambda) & * (t,\lambda) & \sqrt{\lambda} * (t,\lambda) \cr  
\sqrt{\lambda} * (t,\lambda) & \sqrt{\lambda} * (t,\lambda) & 
\lambda * (t,\lambda) \end{array} \right) $$
where all $* (t,\lambda)$ are regular matrix-functions at
$\lambda \rightarrow 0$.

 For the densities of Casimirs (\ref{annlnu}) and
Hamiltonian functions (\ref{h0klnu})-(\ref{h1slnu}) we can 
write now the equations

$${d {\tilde n}^{\nu}(t,\lambda) \over dt} =
\left[ U^{\mu}_{t} {\bf e}_{(\mu)}(t,\lambda)\right]^{\nu} =
\epsilon^{\nu} \langle U^{\mu}_{t} {\bf e}_{(\mu)}(t,\lambda),
{\bf e}_{\nu}(0) \rangle $$

$${d {\tilde h}^{k}_{(0)}(t,\lambda) \over dt} =
- \langle  U^{\mu}_{t} {\bf e}_{(\mu)}(t,\lambda),
{\bf N}_{(0)(k)}(0) \rangle $$

$${d {\tilde h}^{s}_{(1)}(t,\lambda) \over dt} =
- \langle  U^{\mu}_{t} {\bf e}_{(\mu)}(t,\lambda),
{\bf N}_{(1)(s)}(0) \rangle $$ 
along the same curve $\gamma(t)$ on ${\cal M}^{N}$. Since
$\gamma(t)$ is just the arbitrary curve we have that 

$${\tilde n}^{\nu}(U,\lambda) = * (U,\lambda) \,\,\, , \,\,\,
{\tilde h}^{k}_{(0)}(U,\lambda) = * (U,\lambda) \,\,\, , \,\,\,
{\tilde h}^{s}_{(1)}(U,\lambda) = \sqrt{\lambda} * (U,\lambda) $$
where $* (U,\lambda)$ are regular at $\lambda \rightarrow 0$.

 It is easy to see that the expression (\ref{momfer}) 
for the Momentum Operator is 
regular at $\lambda \rightarrow 0$ in this case and we can put

$$h^{k}_{(0)}(U,\lambda) = {\tilde h}^{k}_{(0)}(U,\lambda)
\,\,\, , \,\,\,
h^{s}_{(1)}(U,\lambda) = {1 \over \sqrt{\lambda} }
{\tilde h}^{s}_{(1)}(U,\lambda) $$
to be the regular at $\lambda \rightarrow 0$ densities
of Hamiltonian functions $H_{(0)}^{k}(\lambda)$,
$H_{(1)}^{s}(\lambda)$.

 According to the geometric construction we have

$${\partial h^{k}_{(0)}(U,\lambda) \over \partial U^{\mu} }|_{z}
\equiv 0 \,\,\, , \,\,\,
{\partial h^{s}_{(1)}(U,\lambda) \over \partial U^{\mu} }|_{z}
\equiv 0 $$
and also

$${\partial p(U,\lambda) \over \partial U^{\mu} }|_{z} \equiv 0$$

 Since the functions ${\tilde n}^{\nu}(U,\lambda)$ give locally
the coordinate system on ${\cal M}^{N}$ at every $\lambda$
we have

$$ det \left( {\partial {\tilde n}^{\nu} \over \partial U^{\mu}}
(z,\lambda) \right) \neq 0 $$
and we can put

$$n^{\nu}(U,\lambda) = 
{\partial {\tilde n}^{\nu} \over \partial U^{\xi}}(z,0) \,
{\partial U^{\xi} \over \partial {\tilde n}^{\eta}}(z,\lambda) \,
{\tilde n}^{\eta}(U,\lambda)$$
such that

$${\partial n^{\nu}(U,\lambda) \over \partial U^{\mu}}|_{z} 
\equiv {\partial n^{\nu} \over \partial U^{\mu}} (z,0) =
({\bf e}_{(\mu)})^{\nu} \,\,\, , \,\,\, \nu = 1, \dots, N $$
where ${\bf e}_{(\mu)}$ are the tangent vectors at the point 
$z$ for $\lambda = 0$, $n^{\nu}(U,0) = {\tilde n}^{\nu}(U,0)$. 
So we get parts (I) and (II) of the theorem.

\vspace{0.3cm}

 For the part (III) we prove here first the Lemma:

\vspace{0.3cm}

{\bf Lemma 2.}

{\it Under the conditions formulated in the theorem the functionals

\begin{equation}
\label{npf}
N^{\nu}_{p} = \int_{-\infty}^{+\infty} n^{\nu}_{p}(U) dX 
\,\,\, , \,\,\,
P_{q} = \int_{-\infty}^{+\infty} p_{q}(U) dX 
\end{equation}

\begin{equation}
\label{hhf}
H^{k}_{(0)l} = \int_{-\infty}^{+\infty} h^{k}_{(0)l}(U) dX
\,\,\, , \,\,\,
H^{s}_{(1)t} = \int_{-\infty}^{+\infty} h^{s}_{(1)t}(U) dX 
\end{equation}               
$p,q,l,t = 0, 1, \dots $, generate the local flows 
with respect to both ${\hat J}_{(0)}$ and ${\hat J}_{(1)}$
and commute with the functionals $N^{\mu}_{0}$, $P_{0}$, 
$H^{m}_{(0)0}$ and $H^{n}_{(1)0}$ with respect to the bracket 
${\hat J}_{(0)}$
}

\vspace{0.3cm}

 Proof.

Since the functionals $N^{\mu}_{0}$ are the annihilators of 
${\hat J}_{(0)}$ they commute with all other functionals with
respect to ${\hat J}_{(0)}$ on $L({\cal M}^{N},z)$. Also from
the translational invariance of all functionals 
$N^{\nu}_{p}$, $P_{q}$, $H^{k}_{(0)l}$, $H^{s}_{(1)t}$ we get that
they commute also with the momentum operator $P_{0}$ of 
${\hat J}_{(0)}$ with respect to ${\hat J}_{(0)}$.

 We know then that the functionals $N^{\nu}(\lambda)$,
$P(\lambda)$, $H^{k}_{(0)}(\lambda)$ and $H^{s}_{(1)}(\lambda)$ 
generate the local flows with respect to ${\hat J}_{\lambda}$.
In the case of non-degenerate form $Q_{\lambda}^{ks}$ 
(on ${\cal W}$) this means that they are the conservation laws for 
all the flows ${\tilde w}^{\nu}_{n\eta}(U) U^{\eta}_{X}$
introduced in (\ref{wbasis}). So they are the conservation laws 
for the flows $w^{\nu}_{(0)k\eta}(U) U^{\eta}_{X}$ and 
$w^{\nu}_{(1)s\eta}(U) U^{\eta}_{X}$ and generate the local
flows with respect to ${\hat J}_{(0)}$ and ${\hat J}_{(1)}$.
Now since the flows $w^{\nu}_{(0)m\eta}(U) U^{\eta}_{X}$,
$w^{\nu}_{(1)n\eta}(U) U^{\eta}_{X}$ are generated by the 
functionals $H^{m}_{(0)0}$ and $H^{n}_{(1)0}$ with respect to 
${\hat J}_{(0)}$ we get that all 
$N^{\nu}_{p}$, $P_{q}$, $H^{k}_{(0)l}$, $H^{s}_{(1)t}$
should commute with $H^{m}_{(0)0}$ and $H^{n}_{(1)0}$
with respect to ${\hat J}_{(0)}$ on $L({\cal M}^{N},z)$.

{\hfill Lemma is proved.}

\vspace{0.3cm}

 For the commutativity of all
$N^{\nu}_{p}$, $P_{q}$, $H^{k}_{(0)l}$, $H^{s}_{(1)t}$ with 
respect to both brackets we can now use just the common approach
for the bi-Hamiltonian systems (\cite{magri}) writing

$$\delta F_{q} {\hat J}_{(0)} \delta G_{k} = 
- \delta F_{q-1} {\hat J}_{(1)} \delta G_{k} =
\delta F_{q-1} {\hat J}_{(0)} \delta G_{k+1} = \dots =
\delta F_{0} {\hat J}_{(0)} \delta G_{k+q} = 0 $$
for any two functionals $F_{q}$ and $G_{k}$ from the set
(\ref{npf})-(\ref{hhf}) (the same for the bracket ${\hat J}_{(1)}$).

{\hfill Theorem is proved.}

\vspace{0.5cm}

{\bf Remark.}

Let us point out here that the requirement 
of non-degeneracy of the form $Q_{\lambda}^{ks}$
on ${\cal W}$ is important and in general Theorem 2 is not true
without it. As the example we consider here the Poisson pencil
${\hat J}_{\lambda} = {\hat J}_{(0)} + \lambda {\hat J}_{(1)}$
where

$${\hat J}_{(0)} = \left( \begin{array}{cc}
1 & 0 \cr 0 & - 1 \end{array} \right) {d \over dX} +
\left( \begin{array}{cc}  
U^{1} & U^{2} \cr - U^{2} & - U^{1} \end{array} \right)
\left( \begin{array}{c} 
U^{1}_{X} \cr U^{2}_{X} \end{array} \right) D^{-1}
\left( \begin{array}{cc}
1 & 0 \cr 0 & 1 \end{array} \right) 
\left( \begin{array}{c}
U^{1}_{X} \cr U^{2}_{X} \end{array} \right) + $$

\begin{equation}
\label{exj0}
+ \left( \begin{array}{cc}
1 & 0 \cr 0 & 1 \end{array} \right)
\left( \begin{array}{c}
U^{1}_{X} \cr U^{2}_{X} \end{array} \right) D^{-1}
\left( \begin{array}{cc}
U^{1} & U^{2} \cr - U^{2} & - U^{1} \end{array} \right)
\left( \begin{array}{c}
U^{1}_{X} \cr U^{2}_{X} \end{array} \right) 
\end{equation}
and

$${\hat J}_{(1)} = \left( \begin{array}{cc}
0 & 1 \cr 1 & 0 \end{array} \right) {d \over dX} +
\left( \begin{array}{cc}
0 & 0 \cr 2 U^{1} & 0 \end{array} \right)
\left( \begin{array}{c}
U^{1}_{X} \cr U^{2}_{X} \end{array} \right) D^{-1}
\left( \begin{array}{cc}
1 & 0 \cr 0 & 1 \end{array} \right)
\left( \begin{array}{c}
U^{1}_{X} \cr U^{2}_{X} \end{array} \right) + $$

\begin{equation}
\label{exj1}
+ \left( \begin{array}{cc}
1 & 0 \cr 0 & 1 \end{array} \right)
\left( \begin{array}{c}
U^{1}_{X} \cr U^{2}_{X} \end{array} \right) D^{-1}
\left( \begin{array}{cc}
0 & 0 \cr 2 U^{1} & 0 \end{array} \right)
\left( \begin{array}{c}
U^{1}_{X} \cr U^{2}_{X} \end{array} \right) 
\end{equation}

 We have here the three-dimensional space ${\cal W}$ generated by
the flows

$${\tilde w}^{\nu}_{1\eta} U^{\eta}_{X} = 
\left( \begin{array}{cc}
U^{1} & U^{2} \cr - U^{2} & - U^{1} \end{array} \right)
\left( \begin{array}{c}
U^{1}_{X} \cr U^{2}_{X} \end{array} \right) \,\,\, , \,\,\,
{\tilde w}^{\nu}_{2\eta} U^{\eta}_{X} =
\left( \begin{array}{c}
U^{1}_{X} \cr U^{2}_{X} \end{array} \right) \,\,\, , \,\,\,
{\tilde w}^{\nu}_{3\eta} U^{\eta}_{X} =
\left( \begin{array}{cc}
0 & 0 \cr 2 U^{1} & 0 \end{array} \right)
\left( \begin{array}{c}
U^{1}_{X} \cr U^{2}_{X} \end{array} \right) $$

 Operator ${\hat J}_{\lambda}$ can be written as

$${\hat J}_{\lambda} =
\left( \begin{array}{cc}
1 & \lambda \cr \lambda & - 1 \end{array} \right) {d \over dX} +
\left( \begin{array}{cc}
U^{1} & U^{2} \cr - U^{2} + 2 \lambda U^{1} & - U^{1} 
\end{array} \right)
\left( \begin{array}{c} 
U^{1}_{X} \cr U^{2}_{X} \end{array} \right) D^{-1}
\left( \begin{array}{c}
U^{1}_{X} \cr U^{2}_{X} \end{array} \right) + $$

$$+ \left( \begin{array}{c}
U^{1}_{X} \cr U^{2}_{X} \end{array} \right) D^{-1}
\left( \begin{array}{cc}
U^{1} & U^{2} \cr - U^{2} + 2 \lambda U^{1} & - U^{1}
\end{array} \right)
\left( \begin{array}{c}
U^{1}_{X} \cr U^{2}_{X} \end{array} \right) $$
and the form

$$Q_{\lambda} = \left( \begin{array}{ccc}
0 & 1 & 0 \cr
1 & 0 & \lambda \cr
0 & \lambda & 0 \end{array} \right) $$
is degenerate on ${\cal W}$.

 The metric

$$g^{\nu\mu}_{\lambda}(U) = 
\left( \begin{array}{cc}
1 & \lambda \cr \lambda & - 1 \end{array} \right) $$
here is just the flat metric on $R^{2}$, 
$det \, g^{\nu\mu}_{\lambda} \neq 0$, and it is easy to check the 
equations 

$$g^{\nu\xi}_{\lambda} W^{\mu}_{1\xi}(\lambda) =
g^{\mu\xi}_{\lambda} W^{\nu}_{1\xi}(\lambda) \,\,\, , \,\,\,
g^{\nu\xi}_{\lambda} W^{\mu}_{2\xi} =
g^{\mu\xi}_{\lambda} W^{\nu}_{2\xi} $$
where

$$W^{\mu}_{1\xi}(\lambda) =
\left( \begin{array}{cc}
U^{1} & U^{2} \cr - U^{2} + 2 \lambda U^{1} & - U^{1}
\end{array} \right) \,\,\, , \,\,\,
W^{\mu}_{2\xi}(\lambda) =
\left( \begin{array}{cc}
1 & 0 \cr 0 & 1 \end{array} \right) $$

 Also 

$$\partial_{\nu} W^{\xi}_{1\mu}(\lambda) =
\partial_{\mu} W^{\xi}_{1\nu}(\lambda) \,\,\, , \,\,\,
\partial_{\nu} W^{\xi}_{2\mu} = \partial_{\mu} W^{\xi}_{2\nu} $$
in the flat coordinates $U^{1}$, $U^{2}$ and

$$R^{12}_{12} = W^{1}_{1 1}(\lambda) W^{2}_{2 2} +
W^{1}_{2 1} W^{2}_{1 2}(\lambda) - 
W^{2}_{1 1}(\lambda) W^{1}_{2 2} -
W^{2}_{2 1} W^{1}_{1 2}(\lambda) = $$

$$= W^{1}_{1 1}(\lambda) + W^{2}_{1 2}(\lambda) \equiv 0 $$
so ${\hat J}_{\lambda}$ represents a Poisson bracket of 
Ferapontov type for all $\lambda$.

 Both $g^{\nu\mu}_{(0)}$ and $g^{\nu\mu}_{(1)}$ are
non-degenerate in this case. However, the flow
${\tilde w}^{\nu}_{1\eta} U^{\eta}_{X}$ is not Hamiltonian
with respect to ${\hat J}_{(1)}$ and 
${\tilde w}^{\nu}_{3\eta} U^{\eta}_{X}$ is not Hamiltonian
with respect to ${\hat J}_{(0)}$ as follows from

$$g^{\nu\xi}_{(0)} {\tilde w}^{\mu}_{3\xi} \neq
g^{\mu\xi}_{(0)} {\tilde w}^{\nu}_{3\xi} \,\,\, , \,\,\,
g^{\nu\xi}_{(1)} {\tilde w}^{\mu}_{1\xi} \neq
g^{\mu\xi}_{(1)} {\tilde w}^{\nu}_{1\xi} $$

 It is also easy to check that the flows 
${\tilde w}^{\nu}_{1\eta} U^{\eta}_{X}$ and 
${\tilde w}^{\nu}_{3\eta} U^{\eta}_{X}$ do not commute with each
other.

\section{The Recursion Operator and the Higher Hamiltonian
structures.}
\setcounter{equation}{0}

 Let us use now the symplectic operator (\ref{symomnm}) for
non-degenerate bracket ${\hat J}_{(0)}$ and consider 
the Recursion Operator 
${\hat R}^{\nu}_{\mu} = {\hat J}^{\nu\tau}_{(1)} 
{\hat \Omega}_{(0)\tau\mu}$ under   
the assumptions of Theorem 2. 

 We put $v^{\tau}(U,\lambda) = n^{\tau}(U,\lambda)$,
$\tau = 1,\dots,N$, 
$v^{N+k}(U,\lambda) = h^{k}_{(0)}(U,\lambda)$, 
$k = 1,\dots,g_{0}$, $v^{s}_{0}(U) \equiv v^{s}(U,0)$,
$E^{\tau}_{(0)} = \epsilon^{\tau}$, $\tau = 1,\dots,N$,
$E^{N+k}_{(0)} = e_{(0)k}$, $k = 1,\dots,g_{0}$
(where $n^{\tau}(U,\lambda)$, $h^{k}_{(0)}(U,\lambda)$ are 
the functions from Theorem 2) and write the symplectic form
${\hat \Omega}_{(0)\tau\mu}$ as

\begin{equation}
\label{omform}
{\hat \Omega}_{(0)\tau\mu} = \sum_{k=1}^{N+g_{0}} E^{k}_{(0)}
{\partial v^{k}_{0} \over \partial U^{\tau}} D^{-1}
{\partial v^{k}_{0} \over \partial U^{\mu}} 
\end{equation}

 We can write also the operator ${\hat J}^{\nu\mu}_{(0)}$
as

\begin{equation}
\label{jnmform}
{\hat J}^{\nu\mu}_{(0)} = g^{\nu\mu}_{(0)} {d \over dX}
+ b^{\nu\mu}_{(0)\eta} U^{\eta}_{X} + \sum_{k=1}^{N+g_{0}}
E^{k}_{(0)} \left[{\hat J}^{\nu\tau}_{(0)}
{\partial v_{0}^{k} \over \partial U^{\tau}}\right] D^{-1}
\left[{\hat J}^{\mu\sigma}_{(0)}
{\partial v_{0}^{k} \over \partial U^{\sigma}}\right] 
\end{equation}

 For ${\hat R}^{\nu}_{\mu}$ we have the expression:

$${\hat R}^{\nu}_{\mu} = {\hat J}^{\nu\tau}_{(1)}
{\hat \Omega}_{(0)\tau\mu} =
\sum_{k=1}^{N+g_{0}} E^{k}_{(0)} g^{\nu\tau}_{(1)} \left(
{\partial v^{k}_{0} \over \partial U^{\tau}} \right)_{X} D^{-1}
{\partial v^{k}_{0} \over \partial U^{\mu}} +
\sum_{k=1}^{N+g_{0}} E^{k}_{(0)} g^{\nu\tau}_{(1)}
{\partial v^{k}_{0} \over \partial U^{\tau}}
{\partial v^{k}_{0} \over \partial U^{\mu}} + $$

$$+ \sum_{k=1}^{N+g_{0}} E^{k}_{(0)} b^{\nu\tau}_{(1)\eta} 
U^{\eta}_{X}
{\partial v^{k}_{0} \over \partial U^{\tau}} D^{-1}
{\partial v^{k}_{0} \over \partial U^{\mu}} +
\sum_{s=1}^{g_{1}} \sum_{k=1}^{N+g_{0}} e_{(1)s} E^{k}_{(0)}
w^{\nu}_{(1)s\eta} U^{\eta}_{X} D^{-1}
w^{\tau}_{(1)s\zeta} U^{\zeta}_{X}
{\partial v^{k}_{0} \over \partial U^{\tau}} D^{-1}
{\partial v^{k}_{0} \over \partial U^{\mu}} $$
where

$$w^{\tau}_{(1)s\zeta} U^{\zeta}_{X} 
{\partial v^{k}_{0} \over \partial U^{\tau}} \equiv
\left( Q^{k}_{s} \right)_{X} = {d \over dX} Q^{k}_{s} -
Q^{k}_{s} {d \over dX} $$
for some $Q^{k}_{s}(U)$ according to Theorem 2.

 Let us normalize the functions $Q^{k}_{s}(U)$ such that
$Q^{k}_{s}(z) = 0$ and we have

$$\dots D^{-1} {d \over dX} Q^{k}_{s} \dots = 
\dots Q^{k}_{s} \dots $$
on $L({\cal M}^{N},z)$. 

 Also $d/dX \, D^{-1} \equiv {\hat I}$ on $L({\cal M}^{N},z)$
and

$$\sum_{k=1}^{N+g_{0}} E^{k}_{(0)} g^{\nu\tau}_{(1)} \left(
{\partial v^{k}_{0} \over \partial U^{\tau}} \right)_{X} D^{-1}
{\partial v^{k}_{0} \over \partial U^{\mu}} + 
\sum_{k=1}^{N+g_{0}} E^{k}_{(0)} b^{\nu\tau}_{(1)\eta}
U^{\eta}_{X}
{\partial v^{k}_{0} \over \partial U^{\tau}} D^{-1}
{\partial v^{k}_{0} \over \partial U^{\mu}} + $$

$$+ \sum_{s=1}^{g_{1}} \sum_{k=1}^{N+g_{0}} e_{(1)s} E^{k}_{(0)}
w^{\nu}_{(1)s\eta} U^{\eta}_{X} Q^{k}_{s} D^{-1}
{\partial v^{k}_{0} \over \partial U^{\mu}} =
\sum_{k=1}^{N+g_{0}} E^{k}_{(0)} \left[ {\hat J}^{\nu\tau}_{(1)} 
{\partial v^{k}_{0} \over \partial U^{\tau}} \right] D^{-1}
{\partial v^{k}_{0} \over \partial U^{\mu}} $$

 Using also the relation

$$\sum_{k=1}^{N+g_{0}} E^{k}_{(0)}
{\partial v^{k}_{0} \over \partial U^{\mu}} Q^{k}_{s} =
{\hat \Omega}_{(0)\mu\tau} w^{\tau}_{(1)s\eta} U^{\eta}_{X} =
{\partial h^{s}_{(1)}(U,0) \over \partial U^{\mu}} =
{\partial h^{s}_{(1)0}(U) \over \partial U^{\mu}} $$
on $L({\cal M}^{N},z)$ 
(since $\partial h^{s}_{(1)q}/\partial U^{\mu} \, (z) = 0$, $q \geq 0$)
we can write

$${\hat R}^{\nu}_{\mu} = g^{\nu\tau}_{(1)} g_{(0)\tau\mu} +
\sum_{k=1}^{N+g_{0}} E^{k}_{(0)} \left[ {\hat J}^{\nu\tau}_{(1)}
{\partial v^{k}_{0} \over \partial U^{\tau}} \right] D^{-1}
{\partial v^{k}_{0} \over \partial U^{\mu}} -
\sum_{s=1}^{g_{1}} e_{(1)s} w^{\nu}_{(1)s\eta} U^{\eta}_{X}
D^{-1} {\partial h^{s}_{(1)0} \over \partial U^{\mu}} = $$

$$= V^{\nu}_{\mu} - \sum_{k=1}^{N+g_{0}} E^{k}_{(0)}
\left[ {\hat J}^{\nu\tau}_{(0)}
{\partial v^{k}_{1} \over \partial U^{\tau}} \right] D^{-1}
{\partial v^{k}_{0} \over \partial U^{\mu}} -
\sum_{s=1}^{g_{1}} e_{(1)s}
\left[ {\hat J}^{\nu\tau}_{(0)}
{\partial h^{s}_{(1)0} \over \partial U^{\tau}} \right] D^{-1}
{\partial h^{s}_{(1)0} \over \partial U^{\mu}} $$
where

$$v^{k} (U,\lambda) = \sum_{q=0}^{+\infty} v^{k}_{q}(U) \lambda^{q}
\,\,\, ,\,\,\, 
h^{s}_{(1)} (U,\lambda) = 
\sum_{q=0}^{+\infty} h^{s}_{(1)q}(U) \lambda^{q}$$ 
and $V^{\nu}_{\mu}(U) = g^{\nu\tau}_{(1)}(U) g_{(0)\tau\mu}(U)$.
 
 Let us mention here that according to this definition 
${\hat R}^{\nu}_{\mu}$ will act from the left on the vector-fields
(on ${\cal V}_{0}(z)$) and from the right on the gradients of 
functionals on $L({\cal M}^{N},z)$.

\vspace{0.5cm}

{\bf Theorem 3.}

{\it Let us consider the non-degenerate pencil (\ref{lpencil})
with $det \, g^{\nu\mu}_{(0)} \neq 0$
such that the form $Q_{\lambda}^{ks}$ is also
non-degenerate on ${\cal W}$ for small enough
$\lambda \neq 0$. Then:

\vspace{0.3cm}

 (I) Any power $[{\hat R}^{n}]$, $n \geq 1$ of the recursion 
operator can be written in the form:

$$\left[ {\hat R}^{n} \right]^{\nu}_{\mu} =
\left[ {\hat V}^{n} \right]^{\nu}_{\mu} + 
(-1)^{n} \sum_{k=1}^{N+g_{0}} E^{k}_{(0)} \left( \sum_{s=1}^{n}
\left[ {\hat J}^{\nu\tau}_{(0)} 
{\partial v^{k}_{s} \over \partial U^{\tau}} \right] D^{-1}
{\partial v^{k}_{n-s} \over \partial U^{\mu}} \right) + $$

\begin{equation}
\label{rnnumu}
+ (-1)^{n} \sum_{k=1}^{g_{1}} e_{(1)k} \left( \sum_{s=1}^{n}
\left[ {\hat J}^{\nu\tau}_{(0)}
{\partial h^{k}_{(1),s-1} \over \partial U^{\tau}} \right] D^{-1}
{\partial h^{k}_{(1),n-s} \over \partial U^{\mu}} \right) 
\end{equation}

\vspace{0.3cm}

 (II) The higher Hamiltonian structures 
${\hat J}^{\nu\mu}_{(n)} = [{\hat R}^{n}]^{\nu}_{\xi}
{\hat J}^{\xi\mu}_{(0)}$ can be written on ${\cal F}_{0}(z)$
in the following weakly-nonlocal form:

$${\hat J}^{\nu\mu}_{(n)} = \left[{\hat V}^{n}\right]^{\nu}_{\xi}
g^{\xi\mu}_{(0)} {d \over dX} + 
\left[{\hat V}^{n}\right]^{\nu}_{\xi} b^{\xi\mu}_{(0)\eta}  
U^{\eta}_{X} + $$

$$+ (-1)^{n} \sum_{k=1}^{N+g_{0}} E^{k}_{(0)} \left( 
\sum_{s=1}^{n} \left[ {\hat J}^{\nu\tau}_{(0)}
{\partial v^{k}_{s} \over \partial U^{\tau}} \right]
{\partial v^{k}_{n-s} \over \partial U^{\xi}} g^{\xi\mu}_{(0)}
\right) + $$

$$+ (-1)^{n} \sum_{k=1}^{g_{1}} e_{(1)k} \left(
\sum_{s=1}^{n} \left[ {\hat J}^{\nu\tau}_{(0)}
{\partial h^{k}_{(1),s-1} \over \partial U^{\tau}} \right]
{\partial h^{k}_{(1),n-s} \over \partial U^{\xi}}
g^{\xi\mu}_{(0)} \right) + $$

$$+ (-1)^{n+1} \sum_{k=1}^{N+g_{0}} E^{k}_{(0)} \left(
\sum_{s=1}^{n-1} \left[ {\hat J}^{\nu\tau}_{(0)}
{\partial v^{k}_{s} \over \partial U^{\tau}} \right] D^{-1}
\left[ {\hat J}^{\mu\xi}_{(0)}
{\partial v^{k}_{n-s} \over \partial U^{\xi}} \right] \right) + $$    

\begin{equation}
\label{jnnumu}
+ (-1)^{n+1} \sum_{k=1}^{g_{1}} e_{(1)k} \left(
\sum_{s=1}^{n} \left[ {\hat J}^{\nu\tau}_{(0)}
{\partial h^{k}_{(1),s-1} \over \partial U^{\tau}} \right]
D^{-1} \left[ {\hat J}^{\mu\xi}_{(0)}
{\partial h^{k}_{(1),n-s} \over \partial U^{\xi}} \right]
\right) 
\end{equation}
for $n \geq 2$.

\vspace{0.3cm}

 (III) All the "negative" symplectic forms

$${\hat \Omega}_{(-n)\nu\mu} = {\hat \Omega}_{(0)\nu\mu}
\left[{\hat R}^{n}\right]^{\xi}_{\mu} \,\,\, ,
\,\,\, n \geq 1$$
can be represented on ${\cal V}_{0}(z)$ in the form:

\begin{equation}
\label{omnumu}
{\hat \Omega}_{(-n)\nu\mu} = (-1)^{n} \sum_{k=1}^{N+g_{0}}
E^{k}_{(0)} \left( \sum_{s=0}^{n}
{\partial v^{k}_{s} \over \partial U^{\nu}} D^{-1}
{\partial v^{k}_{n-s} \over \partial U^{\mu}} \right) +
(-1)^{n} \sum_{k=1}^{g_{1}} e_{(1)k} \left(
\sum_{s=1}^{n}
{\partial h^{k}_{(1),s-1} \over \partial U^{\nu}} D^{-1}
{\partial h^{k}_{(1),n-s} \over \partial U^{\mu}} \right) 
\end{equation}
}

\vspace{0.5cm}

 Proof.

(I) We have by induction:

$${\hat R}^{\nu}_{\xi} \left[ {\hat R}^{n}\right]^{\xi}_{\mu} =
\left[ {\hat V}^{n+1} \right]^{\nu}_{\mu} +
(-1)^{n} \sum_{k=1}^{N+g_{0}} E^{k}_{(0)} \left( \sum_{s=1}^{n}
\left[ {\hat R}^{\nu}_{\xi} {\hat J}^{\xi\sigma}_{(0)}
{\partial v^{k}_{s} \over \partial U^{\sigma}} \right] D^{-1}
{\partial v^{k}_{n-s} \over \partial U^{\mu}} \right) + $$

$$+ (-1)^{n} \sum_{k=1}^{g_{1}} e_{(1)k} \left( \sum_{s=1}^{n}
\left[ {\hat R}^{\nu}_{\xi} {\hat J}^{\xi\sigma}_{(0)}
{\partial h^{k}_{(1),s-1} \over \partial U^{\sigma}} \right] D^{-1}    
{\partial h^{k}_{(1),n-s} \over \partial U^{\mu}} \right) + $$

$$+ (-1)^{n} \sum_{q=1}^{N+g_{0}} \sum_{k=1}^{N+g_{0}} E^{q}_{(0)}
\left( \sum_{s=1}^{n} \left[ {\hat J}^{\nu\tau}_{(0)}
{\partial v^{q}_{1} \over \partial U^{\tau}} \right] D^{-1}
P^{qk}_{0s} {\partial v^{k}_{n-s} \over \partial U^{\mu}}
\right) + $$

$$+ (-1)^{n} \sum_{q=1}^{N+g_{0}} \sum_{k=1}^{g_{1}} E^{q}_{(0)}
\left( \sum_{s=1}^{n} \left[ {\hat J}^{\nu\tau}_{(0)}
{\partial v^{q}_{1} \over \partial U^{\tau}} \right] D^{-1}
Q^{qk}_{0s} 
{\partial h^{k}_{(1),n-s} \over \partial U^{\mu}} \right) - $$

$$- \sum_{q=1}^{N+g_{0}} E^{q}_{(0)} 
\left[ {\hat J}^{\nu\tau}_{(0)}
{\partial v^{q}_{1} \over \partial U^{\tau}} \right] D^{-1}
{\partial v^{q}_{0} \over \partial U^{\xi}} 
\left[ {\hat V}^{n} \right]^{\xi}_{\mu} + $$

$$+ (-1)^{n} \sum_{q=1}^{g_{1}} \sum_{k=1}^{N+g_{0}} e_{(1)q}
\left( \sum_{s=1}^{n} \left[ {\hat J}^{\nu\tau}_{(0)}
{\partial h^{q}_{(1),0} \over \partial U^{\tau}} \right] D^{-1}
S^{qk}_{0s} {\partial v^{k}_{n-s} \over \partial U^{\mu}}
\right) + $$

$$+ (-1)^{n} \sum_{q=1}^{g_{1}} \sum_{k=1}^{g_{1}} e_{(1)q}
\left( \sum_{s=1}^{n} \left[ {\hat J}^{\nu\tau}_{(0)}
{\partial h^{q}_{(1),0} \over \partial U^{\tau}} \right] D^{-1}
T^{qk}_{0s}
{\partial h^{k}_{(1),n-s} \over \partial U^{\mu}} \right) - $$

$$- \sum_{q=1}^{g_{1}} e_{(1)q} \left[ {\hat J}^{\nu\tau}_{(0)}
{\partial h^{q}_{(1),0} \over \partial U^{\tau}} \right]
D^{-1} {\partial h^{q}_{(1),0} \over \partial U^{\xi}}
\left[ {\hat V}^{n} \right]^{\xi}_{\mu} $$
where we have

$$E^{k}_{(0)} 
{\partial v^{q}_{0} \over \partial U^{\xi}} \left[
{\hat J}^{\xi\sigma}_{(0)}
{\partial v^{k}_{s} \over \partial U^{\sigma}} \right] \equiv
\left( P^{qk}_{0s} \right)_{X} = {d \over dX} P^{qk}_{0s} -
P^{qk}_{0s} {d \over dX} $$

$$e_{(1)k}
{\partial v^{q}_{0} \over \partial U^{\xi}} \left[
{\hat J}^{\xi\sigma}_{(0)}
{\partial h^{k}_{(1),s} \over \partial U^{\sigma}} \right] \equiv
\left( Q^{qk}_{0s} \right)_{X} = {d \over dX} Q^{qk}_{0s} -
Q^{qk}_{0s} {d \over dX}$$

$$E^{k}_{(0)}
{\partial h^{q}_{(1),0} \over \partial U^{\xi}}
\left[ {\hat J}^{\xi\sigma}_{(0)}
{\partial v^{k}_{s} \over \partial U^{\sigma}} \right] \equiv
\left( S^{qk}_{0s} \right)_{X} = {d \over dX} S^{qk}_{0s} -
S^{qk}_{0s} {d \over dX}$$

$$e_{(1)k}
{\partial h^{q}_{(1),0} \over \partial U^{\xi}} 
\left[ {\hat J}^{\xi\sigma}_{(0)} 
{\partial h^{k}_{(1),s} \over \partial U^{\sigma}} \right] \equiv
\left( T^{qk}_{0s} \right)_{X} = {d \over dX} T^{qk}_{0s} -  
T^{qk}_{0s} {d \over dX}$$
for some functions $P^{qk}_{0s}(U)$, $Q^{qk}_{0s}(U)$,
$S^{qk}_{0s}(U)$, $T^{qk}_{0s}(U)$ according to Theorem 2 and
we use the normalization:

$$P^{qk}_{0s}(z) = Q^{qk}_{0s}(z) = S^{qk}_{0s}(z) = 
T^{qk}_{0s}(z) = 0 $$

 We have from Theorem 2:

$${\partial v^{k}_{s} \over \partial U^{\tau}}(z) = 0 \,\,\, ,
\,\,\, s \geq 1 \,\,\, ({\rm actually} \,\,\,
{\partial h^{k}_{(0),0} \over \partial U^{\tau}}(z) = 0 \,\,\,
{\rm also} )$$

$${\partial h^{k}_{(1),s-1} \over \partial U^{\tau}}(z) = 0 
\,\,\, , \,\,\, s \geq 1 $$
so for $s \geq 1$ we can write

$$\left[ {\hat \Omega}_{(0)\nu\xi} {\hat J}^{\xi\tau}_{(0)}
{\partial v^{k}_{s} \over \partial U^{\tau}} \right] =
{\partial v^{k}_{s} \over \partial U^{\nu}} \,\,\, , \,\,\,
\left[ {\hat \Omega}_{(0)\nu\xi} {\hat J}^{\xi\tau}_{(0)}
{\partial h^{k}_{(1),s-1} \over \partial U^{\tau}} \right] =
{\partial h^{k}_{(1),s-1} \over \partial U^{\nu}} $$
on $L({\cal M}^{N},z)$ and

\begin{equation}
\label{recv}
\left[{\hat R}^{\nu}_{\xi} {\hat J}^{\xi\tau}_{(0)}
{\partial v^{k}_{s} \over \partial U^{\tau}} \right] =
\left[ {\hat J}^{\nu\tau}_{(1)}
{\partial v^{k}_{s} \over \partial U^{\tau}} \right] =
- \left[ {\hat J}^{\nu\tau}_{(0)}
{\partial v^{k}_{s+1} \over \partial U^{\tau}} \right]
\end{equation}

\begin{equation}
\label{rech}
\left[{\hat R}^{\nu}_{\xi} {\hat J}^{\xi\tau}_{(0)}
{\partial h^{k}_{(1),s-1} \over \partial U^{\tau}} \right] =
\left[ {\hat J}^{\nu\tau}_{(1)}
{\partial h^{k}_{(1),s-1} \over \partial U^{\tau}} \right] =
- \left[ {\hat J}^{\nu\tau}_{(0)}
{\partial h^{k}_{(1),s} \over \partial U^{\tau}} \right] 
\end{equation}
according to Theorem 1. 

 Let us now multiply the equalities

$$\left[ {\hat J}^{\xi\tau}_{(1)}
{\partial v^{k}_{s} \over \partial U^{\tau}} \right] =
- \left[ {\hat J}^{\xi\tau}_{(0)} 
{\partial v^{k}_{s+1} \over \partial U^{\tau}} \right] 
\,\,\, , \,\,\,
\left[ {\hat J}^{\xi\tau}_{(1)}
{\partial h^{k}_{(1),s} \over \partial U^{\tau}} \right] =
- \left[ {\hat J}^{\xi\tau}_{(0)}
{\partial h^{k}_{(1),s+1} \over \partial U^{\tau}} \right] $$
by ${\hat \Omega}_{(0)\nu\xi}$ from the left. Since

$${\partial v^{k}_{s+1} \over \partial U^{\tau}}(z) =
{\partial h^{k}_{(1),s+1} \over \partial U^{\tau}}(z) = 0
\,\,\, , \,\,\, s \geq 0 $$
we get again from the Theorem 1:

\begin{equation}
\label{gradvr}
{\partial v^{k}_{s+1} \over \partial U^{\nu}} = -
\left[ {\hat \Omega}_{(0)\nu\xi} {\hat J}^{\xi\tau}_{(1)}
{\partial v^{k}_{s} \over \partial U^{\tau}} \right] = -
\left[ {\partial v^{k}_{s} \over \partial U^{\tau}}
{\hat R}^{\tau}_{\nu} \right] \,\,\, , \,\,\, s \geq 0
\end{equation}

\begin{equation}
\label{gradhr}
{\partial h^{k}_{(1),s+1} \over \partial U^{\nu}} = -
\left[ {\hat \Omega}_{(0)\nu\xi} {\hat J}^{\xi\tau}_{(1)}
{\partial h^{k}_{(1),s} \over \partial U^{\tau}} \right] = - 
\left[ {\partial h^{k}_{(1),s} \over \partial U^{\tau}}
{\hat R}^{\tau}_{\nu} \right] \,\,\, , \,\,\, s \geq 0
\end{equation}
(action from the right).

 Using the relation

$$ \left[ f_{X} D^{-1} \right] = - f_{X} $$
for the action from the right of $D^{-1}$ on any $f(U)$
such that $f(z) = 0$ we get that the last six terms 
in the expression for 
${\hat R}^{\nu}_{\xi} \left[{\hat R}^{n}\right]^{\xi}_{\mu}$ 
can be written as

$$- \sum_{q=1}^{N+g_{0}} E^{q}_{(0)}
\left[ {\hat J}^{\nu\tau}_{(0)}
{\partial v^{q}_{1} \over \partial U^{\tau}} \right] D^{-1}
\left[ {\partial v^{q}_{0} \over \partial U^{\xi}}
\left({\hat R}^{n}\right)^{\xi}_{\mu} \right] -
\sum_{q=1}^{g_{1}} e_{(1)q}
\left[ {\hat J}^{\nu\tau}_{(0)}
{\partial h^{q}_{(1),0} \over \partial U^{\tau}} \right] D^{-1}
\left[ {\partial h^{q}_{(1),0} \over \partial U^{\xi}}
\left({\hat R}^{n}\right)^{\xi}_{\mu} \right] = $$

$$= (-1)^{n+1} \sum_{q=1}^{N+g_{0}} E^{q}_{(0)}
\left[ {\hat J}^{\nu\tau}_{(0)}
{\partial v^{q}_{1} \over \partial U^{\tau}} \right] D^{-1}
{\partial v^{q}_{n} \over \partial U^{\mu}} +
(-1)^{n+1} \sum_{q=1}^{g_{1}} e_{(1)q}
\left[ {\hat J}^{\nu\tau}_{(0)}
{\partial h^{q}_{(1),0} \over \partial U^{\tau}} \right] D^{-1}
{\partial h^{q}_{(1),n} \over \partial U^{\mu}} $$

 Using now the relations (\ref{recv}) and (\ref{rech}) we
get the part (I) of the theorem.

\vspace{0.3cm}

 (II) To avoid much calculations we just write that according to 
Theorem 2 we have the relations

$$E^{k}_{(0)} {\partial v^{q}_{n-s} \over \partial U^{\xi}}
\left[ {\hat J}^{\xi\sigma}_{(0)}
{\partial v^{k}_{0} \over \partial U^{\sigma}} \right] \equiv
\left(P^{qk}_{n-s,0}\right)_{X} = {d \over dX} P^{qk}_{n-s,0} -
P^{qk}_{n-s,0} {d \over dX} $$

$$E^{k}_{(0)} 
{\partial h^{q}_{(1),n-s} \over \partial U^{\xi}}
\left[ {\hat J}^{\xi\sigma}_{(0)}
{\partial v^{k}_{0} \over \partial U^{\sigma}} \right] \equiv
\left(S^{qk}_{n-s,0}\right)_{X} = {d \over dX} S^{qk}_{n-s,0} -
S^{qk}_{n-s,0} {d \over dX} $$
for some $P^{qk}_{n-s,0}(U)$, $S^{qk}_{n-s,0}(U)$,
$P^{qk}_{n-s,0}(z) = 0$, $S^{qk}_{n-s,0}(z) = 0$ and so 
the expression
${\hat J}_{(n)} = {\hat R}^{n} {\hat J}_{(0)}$ can be 
written according on ${\cal F}_{0}(z)$
to (\ref{rnnumu}) and (\ref{jnmform}) and Theorem 1 as

$${\hat J}^{\nu\mu}_{(n)} = ({\rm local \, part \, of}  
{\hat R}^{n})
\times ({\rm local \, part \, of}  {\hat J}_{(0)}) + $$

$$+ (-1)^{n} \sum_{k=1}^{N+g_{0}} E^{k}_{(0)} \left(
\sum_{s=1}^{n} \left[ {\hat J}^{\nu\tau}_{(0)}
{\partial v^{k}_{s} \over \partial U^{\tau}} \right]
{\partial v^{k}_{n-s} \over \partial U^{\xi}}
g^{\xi\mu}_{(0)} \right) + $$

$$+ (-1)^{n} \sum_{k=1}^{g_{1}} e_{(1)k} \left(
\sum_{s=1}^{n} \left[ {\hat J}^{\nu\tau}_{(0)}
{\partial h^{k}_{(1),s-1} \over \partial U^{\tau}} \right]
{\partial h^{k}_{(1),n-s} \over \partial U^{\xi}}
g^{\xi\mu}_{(0)} \right) + $$

$$+ \sum_{q=1}^{N+g_{0}} E^{q}_{(0)} \left[
\left({\hat R}^{n}\right)^{\nu}_{\xi} {\hat J}^{\xi\tau}_{(0)}
{\partial v^{q}_{0} \over \partial U^{\tau}} \right] D^{-1}
\left[ {\hat J}^{\mu\sigma}_{(0)}
{\partial v^{q}_{0} \over \partial U^{\sigma}} \right] - $$

$$- (-1)^{n} \sum_{k=1}^{N+g_{0}} E^{k}_{(0)} \left(
\sum_{s=1}^{n} \left[ {\hat J}^{\nu\tau}_{(0)}
{\partial v^{k}_{s} \over \partial U^{\tau}} \right] D^{-1}
\left[ {\hat J}^{\mu\sigma}_{(0)}
{\partial v^{k}_{n-s} \over \partial U^{\sigma}} \right]
\right) - $$

$$- (-1)^{n} \sum_{k=1}^{g_{1}} e_{(1)k} \left(
\sum_{s=1}^{n} \left[ {\hat J}^{\nu\tau}_{(0)}
{\partial h^{k}_{(1),s-1} \over \partial U^{\tau}} \right] D^{-1}
\left[ {\hat J}^{\mu\sigma}_{(0)}
{\partial h^{k}_{(1),n-s} \over \partial U^{\sigma}} \right]
\right) $$
(since ${\hat J}_{(0)}$ is skew-symmetric it's action from
the right differs by sign from the action from the left).
Using now (\ref{recv}) we get the part (II) of the Theorem.
Let us mention also that it is important that we consider the
space ${\cal F}_{0}(z)$ to use the equality

$$D^{-1} {d \over dX} {\partial v^{k}_{0} \over \partial U^{\xi}}
={\partial v^{k}_{0} \over \partial U^{\xi}} $$
for $k = 1,\dots,N$.

\vspace{0.3cm}

(III) We have

$${\hat \Omega}_{(-n)\nu\mu} = {\hat \Omega}_{(0)\nu\xi}
\left({\hat R}^{n}\right)^{\xi}_{\mu} $$

 Using again the functions $P^{qk}_{0s}(U)$ and
$Q^{qk}_{0s}(U)$ we can write

$${\hat \Omega}_{(-n)\nu\mu} = \sum_{k=1}^{N+g_{0}} E^{k}_{(0)}
{\partial v^{k}_{0} \over \partial U^{\nu}} D^{-1}
\left[ {\partial v^{k}_{0} \over \partial U^{\xi}}
\left({\hat R}^{n}\right)^{\xi}_{\mu} \right] + $$

$$+ (-1)^{n} \sum_{k=1}^{N+g_{0}} E^{k}_{(0)} \left(
\sum_{s=1}^{n} \left[ {\hat \Omega}_{(0)\nu\xi}
{\hat J}^{\xi\tau}_{(0)}
{\partial v^{k}_{s} \over \partial U^{\tau}} \right] D^{-1}
{\partial v^{k}_{n-s} \over \partial U^{\mu}} \right) + $$

$$+ (-1)^{n} \sum_{k=1}^{g_{1}} e_{(1)k} \left(
\sum_{s=1}^{n}\left[ {\hat \Omega}_{(0)\nu\xi}
{\hat J}^{\xi\tau}_{(0)}
{\partial h^{k}_{(1),s-1} \over \partial U^{\tau}} \right] 
D^{-1} {\partial h^{k}_{(1),n-s} \over \partial U^{\mu}} 
\right) $$

 Since

$${\partial v^{k}_{s} \over \partial U^{\tau}}(z) = 0 \,\,\, ,
\,\,\,
{\partial h^{k}_{(1),s-1} \over \partial U^{\tau}} = 0 \,\,\, .
\,\,\, s \geq 1$$
we get the part (III) using Theorem 1 and (\ref{recv}) on 
$L({\cal M}^{N},z)$.

{\hfill Theorem is proved.}

\vspace{0.5cm}

 Let us mention also that if both 
$det \, g^{\nu\mu}_{(0)} \neq 0$ and 
$det \, g^{\nu\mu}_{(1)} \neq 0$ (and the form $Q$ in
non-degenerate on ${\cal W}$) then also the series of "negative"
Hamiltonian operators 
${\hat J}_{(-n)} = {\hat R}^{-n} {\hat J}_{(0)}$ and "positive"
symplectic forms 
${\hat \Omega}_{(n)} = {\hat \Omega}_{(0)} {\hat R}^{-n}$
will be weakly nonlocal.
This situation takes place for example in the Hamiltonian
structures of Whitham systems for KdV, NLS and SG hierarchies.
The local bi-Hamiltonian structure for the averaged KdV
hierarchy was constructed in \cite{dubrnov1} 
(see also \cite{dubrnov2}-\cite{dubrnov3}) using the 
(Dubrovin-Novikov) procedure of averaging of local field-theoretical 
brackets for Gardner-Zakharov-Faddeev and Magri brackets. Both
metrics of the corresponding DN-brackets are non-degenerate
 in this case (and there is no requirements on $Q$). Also
in \cite{dubrnov2}-\cite{dubrnov3} the local bracket for the 
averaged "VG-equation"

$$\varphi_{tt} - \varphi_{xx} + V^{\prime}(\varphi) = 0$$
which is the generalization of SG system
using the same procedure was constructed. In \cite{pavlov2}
all the brackets for averaged KdV, NLS and SG hierarchies
having local (DN) or constant curvature (MF) form 
were enumerated using a nice differential-geometrical approach.
It can be shown that all the pencils represented in 
\cite{dubrnov1}-\cite{dubrnov3}, \cite{pavlov2} actually
satisfy the requirements of Theorems 2 and 3.
The recursion operator approach for the local bi-Hamiltonian
structure for the averaged KdV hierarchy 
(in the diagonal form) was investigated in
\cite{alekspav}, \cite{alekseev} and all the "positive"
operators ${\hat J}_{(n)}$ were explicitly found in
\cite{alekseev} in this case. In 
\cite{malts99u}-\cite{malts99i} the general procedure of
averaging of brackets (\ref{weaknon}) which gives the
weakly non-local Hamiltonian operators for the averaged systems
with Hamiltonian structure (\ref{weaknon}) was constructed.
For many "integrable" systems this method
also gives all the "positive"
weakly nonlocal Poisson brackets of Ferapontov type for the
corresponding Whitham hierarchy. However, we see here that
also the "negative" Hamiltonian operators and Symplectic structures
for the averaged KdV, NLS and SG should be weakly non-local  
according to Theorems 2, 3. We believe that there 
should be a good procedure of averaging of "negative" Hamiltonian
operators for the integrable systems giving
the brackets of Ferapontov Type and the general averaging
procedure for the weakly non-local Symplectic Structures

$${\hat \Omega}_{\nu\mu}(x,y) = \sum_{k=1}^{N} 
C^{(k)}_{\nu\mu}(\varphi, \varphi_{x}, \dots) \delta^{(k)}(x-y) +
\sum_{k,s=1}^{G} d_{ks} 
{\delta H_{(k)} \over \delta \varphi(x)} \nu (x-y)
{\delta H_{(s)} \over \delta \varphi(y)} $$
giving the weakly non-local symplectic structures and the
corresponding F-brackets for the Whitham systems.

 We believe also that the general compatible F-brackets 
should be important for the integration of non-diagonalizable
(bi-Hamiltonian) systems which can not be integrated by Tsarev's
method.
 
 Let us mention also that the requirement of non-degeneracy 
of form $Q$ on ${\cal W}$ is also important in Theorem 3. Such,
it is possible to show that the "negative" and "positive"
Poisson structures corresponding to the pair of operators
(\ref{exj0})-(\ref{exj1}) will not be weakly nonlocal.

\end{document}